\documentclass[equations,notcite,notref]{myown}
\usepackage{amssymb,amsmath,amsthm}
\usepackage{eucal}
\usepackage{accents}
\usepackage{epic}
\usepackage{epsfig}
\usepackage{todonotes} 
\JNMPnumberwithin{equation}{section}

\allowdisplaybreaks

\theoremstyle{definition}

\def\BibTeX{{\rm B\kern-.05em{\sc i\kern-.025em b}\kern-.08em
    T\kern-.1667em\lower.7ex\hbox{E}\kern-.125emX}}

\def\m@th{\mathsurround=0pt}
\mathchardef\bracell="0365 
\def\upbrall{$\m@th\bracell$}
\def\undertilde#1{\mathop{\vtop{\ialign{##\crcr
    $\hfil\displaystyle{#1}\hfil$\crcr
     \noalign
     {\kern1.5pt\nointerlineskip}
     \upbrall\crcr\noalign{\kern1pt
   }}}}\limits}   

\newcommand{\cof}{\mathrm{cof}}

\newcommand{\kp}{\kappa}
\newcommand{\ld}{\lambda}
\newcommand{\oa}{\omega}
\newcommand{\bea}{\begin{eqnarray}}
\newcommand{\eea}{\end{eqnarray}}
\newcommand{\bse}{\begin{subequations}}
\newcommand{\ese}{\end{subequations}}
\newcommand{\nn}{\nonumber}
\newcommand{\vf}{\varphi}
\newcommand{\vr}{\varrho}

\newcommand{\wh}{\widehat}

\newcommand{\wt}{\widetilde}
\newcommand{\ut}{\undertilde}

\newcommand{\tbL}{\,^{t\!}{\boldsymbol{L}}}

\newcommand{\tLd}{\,^{t\!}\boldsymbol{\Lambda}}
\newcommand{\btL}{\,^{t\!}\boldsymbol{L}}
\newcommand{\bL}{\boldsymbol{L}}

\newcommand{\bO}{\boldsymbol{O}}
\newcommand{\bA}{\boldsymbol{A}}
\newcommand{\bLd}{\boldsymbol{\Lambda}}
\newcommand{\bB}{\boldsymbol{B}}
\newcommand{\bG}{\boldsymbol{G}}
\newcommand{\bF}{\boldsymbol{F}}

\newcommand{\bR}{\boldsymbol{R}}
\newcommand{\bQ}{\boldsymbol{Q}}
\newcommand{\bV}{\boldsymbol{V}}
\newcommand{\bW}{\boldsymbol{W}}
\newcommand{\bZ}{\boldsymbol{Z}}
\newcommand{\tee}{\,^t\!\boldsymbol{e}}
\newcommand{\bee}{\boldsymbol{e}}
\newcommand{\bU}{\boldsymbol{U}}
\newcommand{\bC}{\boldsymbol{C}}
\newcommand{\bOm}{\pmb{\Omega}}

\newcommand{\buk}{\boldsymbol{u}_k}

\def\hypotilde#1#2{\vrule depth #1 pt width 0pt{\smash{{\mathop{#2}
\limits_{\displaystyle\widetilde{}}}}}}

\newcommand{\bvf}{\boldsymbol{\varphi}} 
\newcommand{\bM}{\boldsymbol{M}}
\newcommand{\bS}{\boldsymbol{S}}

\newcommand{\bu}{{\boldsymbol u}}

 \newcommand{\pl}{\partial}

 \newcommand{\Ld}{\pmb{\Lambda}}

 \newcommand{\bP}{\boldsymbol{P}}
 \newcommand{\tbP}{\,^{t\!}\boldsymbol{P}}

\newcommand{\bE}{{\boldsymbol E}}

\newcommand{\bc}{{\boldsymbol c}}
\newcommand{\tc}{\,^{t\!}\boldsymbol{c}}

\newcommand{\bsg}{{\boldsymbol \sigma}}

 \def\hypotilde#1#2{\vrule depth #1 pt width 0pt{\smash{{\mathop{#2}
 \limits_{\displaystyle\widetilde{}}}}}}

\DeclareMathAccent{\wtilde}{\mathord}{largesymbols}{"65}
\DeclareMathAccent{\what}{\mathord}{largesymbols}{"62}

\makeatletter
\def\wb{\accentset{{\cc@style\underline{\mskip10mu}}}}
\makeatother 

\begin{document}

%  Headings
%
\renewcommand{\evenhead}{F W Nijhoff, C Zhang and D-J Zhang}
\renewcommand{\oddhead}{The elliptic lattice 
KdV system} 

%  Titlepage
%
\thispagestyle{empty}

\FirstPageHead{*}{*}{20**}{\pageref{firstpage}--\pageref{lastpage}}{Article}
%  Parameters: Volume, number, year, page range, paper type
%  'Article' could be changed to 'Letter' or 'Review Article'

%\copyrightnote{2002}{F W Nijhoff}

\Name{The elliptic lattice KdV system revisited} 

\label{firstpage}

%\Date{Received May *, 2002; Revised Month *, 200*; 
%Accepted Month *, 200*}
\Author{Frank W NIJHOFF\,$^a$, Cheng ZHANG\,$^b$ and Da-jun ZHANG\,$^c$}

\Address{$^a$School of Mathematics, University of 
Leeds, Leeds LS2 9JT, United Kingdom \\
~~E-mail: frank.nijhoff@gmail.com\\[10pt]
$^b$ Department of Mathematics, Shanghai University, 
Shanghai 200444, PR China\\
~~E-mail: ch.zhang.maths@gmail.com\\[10pt]
$^c$ Department of Mathematics, Shanghai University, 
Shanghai 200444, PR China\\
~~E-mail: djzhang@shu.edu.cn}

\begin{abstract}
\noindent 
In a previous paper \cite{NijPutt}, a $2$-parameter extension of the lattice potential 
KdV equation was derived, associated with an elliptic curve. This comprises a rather 
complicated $3$-component system on the quad lattice which contains the moduli of the 
elliptic curve as parameters. In the present 
paper, we investigate this system further and, among other results, we 
derive a $2$-component multiquartic form of the system on the quad lattice. 
Furthermore, we construct an elliptic Yang-Baxter map, and study the 
associated continuous and semi-discrete systems. In particular, 
we derive the so-called ``generating PDE" for this system, comprising 
a $6$-component system of second order PDEs which could be considered to constitute 
an elliptic extension of the Ernst equations of General Relativity. 
\end{abstract}

%  The paper
%
\section{Introduction}

The elliptic difference-difference Korteweg-de Vries (Ell$\Delta\Delta$KdV) system was introduced 
in \cite{NijPutt}. It is given by a three-component system of the form 
\bse\label{eq:EllDDKdV}\begin{eqnarray}
&&\left(p+q+u-\wh{\wt{u}}\right)\left(p-q+\wh{u}-\wt{u}\right)
=p^2-q^2+g\left(\wt{s}-\wh{s}\right)\left(\wh{\wt{s}}-s\right)\ , \label{eq:EllDDKdVa}\\
&& \left[ \left(p+u-\frac{\wt{w}}{\wt{s}}\right)\wt{s}-\left(q+u-\frac{\wh{w}}{\wh{s}}
\right)\wh{s}\right]\wh{\wt{s}} \nn \\
&& ~~~~~~~~~~
= \left[ \left(p-\wh{\wt{u}}+\frac{\wh{w}}{\wh{s}}\right)\wh{s}
-\left(q-\wh{\wt{u}}+\frac{\wt{w}}{\wt{s}}\right)\wt{s}\right]s\ , \label{eq:EllDDKdVb}\\ 
&& \left[ \left(p-\wt{u}+\frac{w}{s}\right)s+\left(q+\wt{u}-
\frac{\wh{\wt{w}}}{\wh{\wt{s}}}\right)\wh{\wt{s}}\right]\wh{s} \nn \\
&& ~~~~~~~~~~
= \left[ \left(p+\wh{u}-\frac{\wh{\wt{w}}}{\wh{\wt{s}}}\right)\wh{\wt{s}}+
\left(q-\wh{u}+\frac{w}{s}\right)s\right]\wt{s}\ , \label{eq:EllDDKdVc}\\ 
&& \left(p+u-\frac{\wt{w}}{\wt{s}}\right)\left(p-\wt{u}+
\frac{w}{s}\right)=p^2-R(s\wt{s})\ , \label{eq:EllDDKdVd}\\ 
&& \left(q+u-\frac{\wh{w}}{\wh{s}}\right)\left(q-\wh{u}+
\frac{w}{s}\right)=q^2-R(s\wh{s})\ ,  \label{eq:EllDDKdVe}
\end{eqnarray} \ese
for three functions $u(n,m)$, $s(n,m)$ and $w(n,m)$ of discrete (lattice) 
variables $n,m$ that are 
linked to the lattice parameters $p$ and $q$ respectively. 
We have employed here the abbreviated notation for lattice shifts, see 
\cite{HJN}, i.e. for functions $f=f(n,m)$, we denote: 
\[ \wt{f}=f(n+1,m)\ , \quad 
\wh{f}=f(n,m+1)\ , \quad \wh{\wt{f}}=f(n+1,m+1)\ .  \]   
The system \eqref{eq:EllDDKdV} contains an elliptic curve: 
\begin{equation} \Gamma:\quad x^2=R(1/X)=X+3e+g/X\ ,  
\label{eq:ellcurve} 
\end{equation} 
which is the standard Weierstrass curve in disguise; in fact it can be parametrised as follows in terms of Weierstrass functions 
\[ x=\tfrac{1}{2}\frac{\wp'(\xi)}{\wp(\xi)-e}\ , \quad 
X=\wp(\xi)-e\  , \]
where $\wp(\xi)$ is the usual Weierstrass function, and where the 
parameter $e$ is one of the branch points of the Weierstrass curve (which we fix),  
and $g=(e-e')(e-e'')$ with $e'$ and $e''$ the other branch points. 

Note that in the limit $g\to 0$, i.e. when the curve degenerates, in 
\eqref{eq:EllDDKdVa}  the terms involving $s$ disappear, and effectively that equation 
reduces to the lattice potential KdV equation (H1 in the ABS list of \cite{ABS}), and 
decouples from the remaining equations in the system. Thus, the system 
\eqref{eq:EllDDKdV} can be considered to constitute an elliptic extension of the H1 
equation. In that respect the Ell$\Delta\Delta$KdV system can be compared to the 
well-known Q4, (otherwise known as Adler's equation) 
in the ABS list that was discovered by Adler in 1997 \cite{Adler}, as the permutability 
condition for the well-known Krichever-Novikov equation. Both the Ell$\Delta\Delta$KdV 
system and the Q4 equation can be considered to be examples of \textit{elliptic discrete 
integrable systems}, which comprises also the lattice Landau-Lifschitz equation \cite{LL}, by 
virtue of the fact that they are defined in terms of an elliptic curve (i.e. 
contain the modulo of an elliptic curve as parameters of the equation itself) and 
not merely are equations that possess elliptic solutions (which is true for many 
soliton equations). 

In \cite{NijPutt}, where the system \eqref{eq:EllDDKdV} was constructed using  
a formalism of what was called an algebra of \textit{elliptic matrices}, together 
with an associated direct linearisation scheme based on an elliptic Cauchy kernel, 
already a number of key properties of the system  
were established: a Lax pair, elliptic multi-soliton solutions and the 
consistency of the initial value problems on the two-dimensional lattice. 
Later, the system \eqref{eq:EllDDKdV} together with its solutions was reconstructed in 
\cite{SunZhangNij2017} by using the Cauchy matrix approach. 
Furthermore, by construction the system is multdimensionally consistent (MDC) 
in the sense of \cite{NW,BS}, also known as consistency-around-the-cube (CAC), cf.  \cite{HJN}. In the present paper we develop the elliptic KdV system further: 
we derive a $2$-component multi-quartic form for the system, and derive an 
elliptic Yang-Baxter map. Furthermore, we investigate the relevant $\tau$-function 
which necessitates an extension of the system with a number of parameter-dependent 
functions which obey additional systems of equation. In fact, a 2$\times$2 matrix 
system emerges from the analysis which leads to a matrix Schwarzian KdV subject to 
a constraint arising from the elliptic curve. Additionally, we investigate 
associated continuous and semi-discrete systems, one of which can be considered 
to be an elliptic Toda type system, while the other is an elliptic extension of 
the dressing chain. Finally we present a fully continuous coupled system of six
second order PDE's, which is the elliptic extension of the \textit{generating PDE} 
introduced in \cite{NHJ} associated with the conventional KdV hierarchy, and 
which can be thought of as an elliptic extension of the Ernst equations of 
General Relativity.

\section{Applications of the elliptic KdV system}
\setcounter{equation}{0}

In this section we present a few structural results and direct consequences of the elliptic KdV 
system \eqref{eq:EllDDKdV}. 

\subsection{Lax pair and gauge transformation -- dual system} 

The Lax pair for the lattice system \eqref{eq:EllDDKdV} was derived in 
\cite{NijPutt}, and it gives the elliptic KdV system via the zero curvature 
condition: 
\begin{equation}\label{eq:zerocurv} 
\wt{\bvf}=\bL(K)\bvf\  , \quad  \wh{\bvf}=\bM(K)\bvf\quad \Rightarrow\quad \wt{\bL} \bM=\wh{\bM} \bL
\end{equation} 
with matrices $\bL(K)$ and $\bM(K)$, depending on a spectral 
parameter $K$, given by:
\bse\label{eq:LaxLM}\begin{equation}\label{eq:LaxL} 
\bL(K)=\left( \begin{array}{cc}
p-\wt{u}+\frac{g}{K}\wt{s}w & 1-\frac{g}{K}\wt{s}s \\
\begin{array}{l} K+3e-p^2+g\wt{s}s \\
+(p-\wt{u})(p+u)+\frac{g}{K}\wt{w}w \end{array} & p+u-\frac{g}{K}\wt{w}s
\end{array}\right)
\end{equation}
and
\begin{equation}\label{eq:LaxM}
\bM(K)=\left( \begin{array}{cc}
q-\wh{u}+\frac{g}{K}\wh{s}w & 1-\frac{g}{K}\wh{s}s \\
\begin{array}{l} K+3e-q^2+g\wh{s}s \\
+(q-\wh{u})(q+u)+\frac{g}{K}\wh{w}w \end{array} & q+u-\frac{g}{K}\wh{w}s
\end{array}\right)\ . 
\end{equation} \ese 

Note that these matrices have determinant
\begin{align} \label{eq:LaxDet}
&  \det(\bL(K))= p^2-k^2 +\frac{g}{K} s\wt{s}\,\Gamma_p\  , \quad {\rm where} \\ 
& \Gamma_p=\left(p+u-\wt{w}/\wt{s}\right)\left(p-\wt{u}+w/s\right) 
-p^2+R(s\wt{s})\  ,  \nonumber 
\end{align} 
and similar for $\bM(K)$, with $(k, K)$ on the elliptic curve $\Gamma$. Note that the Lax equations require $\Gamma_p=\Gamma_q=0$, in which case we have $\det(\bL(K))=p^2-k^2$.  
%The discrete Lax equation
%\[ \wt{\bL} \bM=\wh{\bM} \bL  ~~~\Rightarrow~~~ {\rm Elliptic\ \ Lattice\ \ KdV\ \ System} \]

Eliminating variables from the set of relations in subsection 3.5 (below) 
choosing variables in an alternative way, we can 
derive a  dual system in terms of $s$,$v$ and $h$:
\bse\label{eq:dualsyst}\begin{align}
&\left(p+q+g\wh{\wt{h}}-gh\right)\left(p-q+g\wt{h}-g\wh{h}\right) 
=p^2-q^2+g\left(\wt{s}-\wh{s}\right)\left(\wh{\wt{s}}-s\right)\ , \label{eq:dualsysta}  \\ 
& \left[\left(p-gh+ \frac{\wt{v}}{\wt{s}}\right)\wt{s}-
\left(q-gh+\frac{\wh{v}}{\wh{s}}\right)\wh{s}\right]\wh{\wt{s}} =
\left[ \left(p+g\wh{\wt{h}}-\frac{\wh{v}}{\wh{s}}\right)\wh{s}
-\left(q+g\wh{\wt{h}} - \frac{\wt{v}}{\wt{s}}\right)\wt{s}\right]s\ , \label{eq:dualsystb} \\ 
& \left[ \left(p+g\wt{h}-\frac{v}{s}\right)s
+\left(q-g\wt{h}+\frac{\wh{\wt{v}}}{\wh{\wt{s}}}\right)\wh{\wt{s}}\right]\wh{s}  = \left[ \left(p-g\wh{h}+\frac{\wh{\wt{v}}}{\wh{\wt{s}}}\right)\wh{\wt{s}}
+\left(q+g\wh{h}-\frac{v}{s}\right)s\right]\wt{s}\ , \label{eq:dualsystc} \\ 
& \left(p-gh+\frac{\wt{v}}{\wt{s}}\right)\left(p+g\wt{h}-
\frac{v}{s}\right)=p^2-R(s\wt{s})\ , \label{eq:dualsystd} \\
& \left(q-gh+\frac{\wh{v}}{\wh{s}}\right)\left(q+g\wh{h}-
\frac{v}{s}\right)=q^2-R(s\wh{s})\ ,  \label{eq:dualsyste}
%\end{center} 
\end{align}\ese  
which is Miura related to the $u,w,s$ system. In fact the Lax pair for the dual 
system: 
\begin{align}  
& \wt{\vr}=\mathcal{L}(K)\vr\ , \quad \wh{\vr}=\mathcal{M}(K)\vr \label{eq:Laxb} 
\end{align} 
with Lax matrices 
\begin{equation} \label{eq:dualLax}
\mathcal{L}(K)=\left( \begin{array}{cc} 
p-gh+K\wt{v}s & 
\begin{array}{l} \frac{g}{K}+3e-p^2+g\wt{s}s \\ 
+(p+g\wt{h})(p-gh)+K\wt{v}v \end{array} \\ 
1-K\wt{s}s & p+g\wt{h}-K\wt{s}v 
\end{array}\right) 
\end{equation}
(and similarly for $\mathcal{M}$ replacing $p\to q$, $\wt{\phantom{a}}\to \wh{\phantom{a}}$), 
which are gauge related to the matrices $L(K)$ and $M(K)$ via: 
\begin{equation}\label{eq:gauge}  
\vf=K\boldsymbol{G}\vr\quad ,\quad\boldsymbol{G}=
\left( \begin{array}{cc} s & v \\ w & \frac{vw-1}{s}\end{array}\right) \quad {\rm and}\quad 
\wt{\boldsymbol{G}}\mathcal{L}=L\boldsymbol{G}\  .
\end{equation}  
 In fact, \eqref{eq:gauge} encodes the Miura relations which form part of the 
 set of relations in subsection 3.5, cf. also \cite{NijPutt}.

\subsection{Multiquartic system} 

Solving the relations \eqref{eq:EllDDKdVd} and \eqref{eq:EllDDKdVe}, which are 
quadratic in $p+u-\wt{u}$ and $q+u-\wh{u}$ respectively after rewriting them 
in terms of $U=u-w/s$, we get 
\begin{equation}\label{eq:uUcorresp} 
p+u-\wt{u}=\tfrac{1}{2}(U-\wt{U})+D_p\  , \quad 
q+u-\wh{u}=\tfrac{1}{2}(U-\wh{U})+D_q\  , 
\end{equation} 
in which  the discriminants $D_p$ and $D_q$ are given (up to a sign) by 
\be\label{eq:DpDq} 
D_p^2:= \tfrac{1}{4}\left(U+\wt{U}\right)^2+p^2-R(s\wt{s})\ , 
\quad D_q^2:= \tfrac{1}{4}\left(U+\wh{U}\right)^2+q^2-R(s\wh{s})\ . \ee 
From \eqref{eq:uUcorresp} we have  
\bse\begin{align} 
& D_p-D_q=\wh{D}_p-\wt{D}_q= p-q+\wh{u}-\wt{u}+\tfrac{1}{2}(\wt{U}-\wh{U})\ ,  \\ 
& D_p+\wt{D}_q=D_q+\wh{D}_p= p+q+u-\wh{\wt{u}}+\tfrac{1}{2}(\wh{\wt{U}}-U)\ ,   
\end{align}\ese 
Thus, from \eqref{eq:EllDDKdVa} and respectively the sum of \eqref{eq:EllDDKdVb} and \eqref{eq:EllDDKdVc} we get 
\bse \begin{align}\label{eq:1steq} 
& (U-\wh{\wt{U}})(D_p-D_q)+(\wh{U}-\wt{U})(D_p+\wt{D}_q)=(\wt{s}-\wh{s})(\wh{\wt{s}}-s)\left(g-\frac{1}{s\,\wt{s}\,\wh{s}\,\wh{\wt{s}}}\right) \ ,   \\ 
& \frac{\wh{s}+\wt{s}}{\wh{s}-\wt{s}}(D_p-D_q)- 
\frac{\wh{\wt{s}}+s}{\wh{\wt{s}}-s}(D_p+\wt{D}_q)=\tfrac{1}{2}(U+\wh{U}+\wt{U}+\wh{\wt{U}})\  , \label{eq:2ndeq}
\end{align} 
while the difference between \eqref{eq:EllDDKdVb} and \eqref{eq:EllDDKdVc} yields 
\begin{equation}
D_p -\wh{D}_p=D_q-\wt{D}_q=
\tfrac{1}{2}\frac{\wh{\wt{s}}+s}{\wh{\wt{s}}-s}(\wh{\wt{U}}-U) 
-\tfrac{1}{2} \frac{\wh{s}+\wt{s}}{\wh{s}-\wt{s}}(\wh{U}-\wt{U})=:\tfrac{1}{2}\Delta\ ,   
\label{eq:3deq} 
\end{equation}\ese  
where $\Delta$ is the determinant of the linear system \eqref{eq:1steq}, \eqref{eq:2ndeq} 
for $D_p-D_q$ and $D_p+\wt{D}_q$. 
%and noting that the determinant $\Gamma$, given by 
%\[ \Gamma=\frac{\wt{s}+\wh{s}}{\wt{s}-\wh{s}}(\wt{U}-\wh{U})
%+\frac{\wh{\wt{s}}+s}{\wh{\wt{s}}-s}(U-\wh{\wt{U}})\ , \]
%equals the r.h.s. of \eqref{eq:3deq}, 
Solving the latter and using \eqref{eq:1steq} we get the following expressions for $D_p$ and $D_q$: 
\bse \label{eq:DpDqexpl}
\begin{align}\label{eq:Dp} 
D_p &= \tfrac{1}{4}\Delta+\frac{1}{\Delta}\left[  
(\wh{s}\wh{\wt{s}}-s\wt{s})\left(g-\frac{1}{s\wh{s}\wt{s}\wh{\wt{s}}}\right) 
+\tfrac{1}{4}(U+\wt{U})^2-\tfrac{1}{4}(\wh{U}+\wh{\wt{U}})^2\right]\ , \\ 
\wh{D}_p &= -\tfrac{1}{4}\Delta+\frac{1}{\Delta}\left[  
(\wh{s}\wh{\wt{s}}-s\wt{s})\left(g-\frac{1}{s\wh{s}\wt{s}\wh{\wt{s}}}\right) 
+\tfrac{1}{4}(U+\wt{U})^2-\tfrac{1}{4}(\wh{U}+\wh{\wt{U}})^2\right]\ , 
%D_q &= -\tfrac{1}{4}\Gamma+\frac{1}{\Gamma}\left[  
%(s\wh{s}-\wt{s}\wh{\wt{s}})\left(g-\frac{1}{s\wh{s}\wt{s}\wh{\wt{s}}}\right) 
%+\tfrac{1}{4}(\wt{U}+\wh{\wt{U}})^2-\tfrac{1}{4}(U+\wh{U})^2\right]\ . 
\label{eq:Dq} 
\end{align}\ese   
(and similar expressions for $D_q$ and $\wt{D}_q$ with $p$ and $q$, and $\wt{\phantom{a}}$ and $\wh{\phantom{a}}$ interchanged). 
Squaring these expressions and equating them to the definitions of 
$D_p^2$ and $D_q^2$, namely \eqref{eq:DpDq}, 
%\[ D_p^2= \tfrac{1}{4}\left(U+\wt{U}\right)^2+p^2-R(s\wt{s})\ , 
%\quad D_q^2= \tfrac{1}{4}\left(U+\wh{U}\right)^2+q^2-R(s\wh{s})\ , \]
we get a coupled system of multi-quartic quadrilateral 
equations for $s$ and $U$.

% $\mho$ $\mathbf{\mho}$

\subsection{Elliptic Yang-Baxter map}

Consider variables $X=1/(s\wt{s})$ and $Y=1/(s\wh{s})$ and set 
\[ x^2=R(s\wt{s})=X+3e+g/X\ ,\quad y^2=R(s\wh{s})=Y+3e+g/Y\ , \] 
implying $(x,X)$ and $(y,Y)$ lie on the elliptic curve. 
Let $\omega$ be a half-period, with $e=\wp(\omega)$, and 
parametrising the curve with uniformising variables $\xi,\eta$, 
we can identify: 
\begin{align*} 
& x=\zeta(\xi+\oa)-\zeta(\xi)-\zeta(\oa)\quad \Rightarrow \quad 
X=\frac{1}{s\wt{s}}=\wp(\xi)-e\ \ {\rm and}\ \ gs\wt{s}=\wp(\xi+\oa)-e\ , \\ 
& y=\zeta(\eta+\oa)-\zeta(\eta)-\zeta(\oa)\quad \Rightarrow\quad  
Y=\frac{1}{s\wh{s}}=\wp(\eta)-e\ \ {\rm and}\ \ gs\wh{s}=\wp(\eta+\oa)-e\ . 
\end{align*}  
From the Ell$\Delta\Delta$KdV we then have $aA=p^2-x^2$ and 
$bB=q^2-y^2$, where 
\[ a\equiv p+u-\frac{\wt{w}}{\wt{s}}, \quad A=p-\wt{u}+\frac{w}{s}\ , 
\quad b\equiv q+u-\frac{\wh{w}}{\wh{s}}, \quad B=q-\wh{u}+\frac{w}{s}\ , \] 
from which we have the identities $\wh{a}-\wt{b}=A-B$~ and 
$a-b=\wh{A}-\wt{B}$. 
Thus, we get 
\bse\begin{equation} \label{eq:YBI} 
\wh{a}-\wt{b}=\frac{p^2-x^2}{a}-\frac{q^2-y^2}{b}\ , \quad 
a-b=\frac{p^2-\wh{x}^2}{\wh{a}}-\frac{q^2-\wt{y}^2}{\wt{b}}\  . 
\end{equation} 
Furthermore from the intermediate relations in Ell$\Delta\Delta$KdV 
we get 
\begin{equation} \label{eq:YBII}
\frac{a}{\wt{Y}}-\frac{b}{\wh{X}}=\frac{p^2-\wh{x}^2}{\wh{a}Y} -
\frac{q^2-\wt{y}^2}{\wt{b}X}\ , \quad 
\frac{\wh{a}}{\wt{Y}}-\frac{\wt{b}}{\wh{X}}= \frac{p^2-x^2}{aY} -\frac{q^2-y^2}{bX}\  , 
\end{equation} 
together with 
\begin{equation} \label{eq:YBIII}
\wh{a}b-\wt{b}a=\wh{X}-\wt{Y}+g\left(\frac{1}{Y}-\frac{1}{X}\right)\ . 
\end{equation}\ese

{\bf Conjecture:} The map $\left( (a,(x,X));(\wt{b},(\wt{y},\wt{Y}))\right)\stackrel{R_{pq}}{\rightarrow} \left( (\wh{a},(\wh{x},\wh{X}));(b,(y,Y))\right)$ 
given by \eqref{eq:YBI}, \eqref{eq:YBII}, \eqref{eq:YBIII}, is an (elliptic) Yang-Baxter map, 
with $(x,X)$ and $(y,Y)$ 
on the elliptic curve, and subject to $X\wh{X}=Y\wt{Y}$.  

%\begin{columns}[t]

%\begin{column}{5cm}

\begin{figure}[ht]
\centering
%\centering
%Schematically:
%\vspace{.2cm}
\setlength{\unitlength}{.7mm}
\begin{picture}(40,40)(0,0)

\put(0,0){\circle*{3}}
\put(0,30){\circle*{3}}
\put(30,30){\circle*{3}}
\put(30,0){\circle*{3}}

\put(0,0){\vector(1,0){15}}
\put(15,0){\vector(1,0){15}}
\put(0,0){\vector(0,1){15}}
\put(0,15){\vector(0,1){15}}
\put(0,30){\vector(1,0){15}}
\put(15,30){\vector(1,0){15}}
\put(30,0){\vector(0,1){15}}
\put(30,15){\vector(0,1){15}}

\put(16,15){$R_{p,q}$} 
\put(25,5){\vector(-1,1){20}}

\begin{color}{blue}
\put(-7,-7){$(u,s,w)$}
\put(33,-4){$(\wt{u},\wt{s},\wt{w})$}
\put(33,33){$(\wh{\wt{u}},\wh{\wt{s}},\wh{\wt{w}})$}
\put(-5,34){$(\wh{u},\wh{s},\wh{w})$}

\put(15,2){$\mathcal{A}$}
\put(15,33){$\wh{\mathcal{A}}$}
\put(-5,16){$\mathcal{B}$}
\put(35,16){$\wt{\mathcal{B}}$}
\end{color}

%\put(80,30){ Here $p$, $q$ are \textit{lattice parameters}}
%\put(120,20){ $\bv~~\stackrel{\begin{color}{red}p\end{color}}{\rightarrow}~~ \wt{v}$}
%\put(80,15){lattice shifts:}
%\put(120,10){ $\bv~~\stackrel{\begin{color}{red}q\end{color}}{\rightarrow}~~ \wh{\bv}$}

\end{picture}
\end{figure}
\vspace{.5cm} 

%\end{column}

%\begin{column}{7cm}
%\centering

\noindent 
To prove the conjecture, which seems to be true by its construction, 
we need to assert the properties of YB map:  
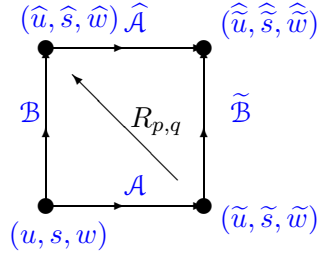
\begin{description} 
\item{{\bf i)} Yang-Baxter relation:}  
\[ R_{p,q}^{(1,2)}\circ R_{p,r}^{(1,3)}\circ R_{q,r}^{(2,3)}=
R_{q,r}^{(2,3)}\circ R_{p,r}^{(1,3)}\circ R_{p,q}^{(1,2)}\ , \]  
\item{{\bf ii)} inversion relation:}
\[ R_{p,q}\circ R^{\ast}_{p,q}={\rm id}\ , \] 
where the map $R^\ast_{p,q}$ is a conjugate YB map. 
\end{description} 
The conjugate map $R^\ast_{p,q}$ arises from the inversion symmetry: 
\[ p\to -p,~ q\to -q, (x,X)\to (x,X)\  , {\rm and}\quad a\to -(p^2-x^2)/a,~b\to -(q^2-y^2)/b\ . \]  
%\end{column}
A direct proof of the Yang-Baxter relation is still absent, due to the 
highly implicit nature of the defining relations of the map. Finding a 
refactorisation Lax pair for the map, based on the Lax pair 
\eqref{eq:zerocurv} for the EllKdV 
system could settle this problem, but a technical difficulty is the 
absence of a factorisation property for the Lax matrices 
\eqref{eq:LaxL}, \eqref{eq:LaxM}, unlike in the case of H1 equation (see \cite{PapVes}, 
and also subsection 3.4 of \cite{HJN}).

\subsection{Elliptic Toda type system: EllD$\Delta$KdV}

Through a skew continuum limit (cf. Ch. 5 of \cite{HJN}), we obtain from eqs. (\ref{eq:EllDDKdVa})-\eqref{eq:EllDDKdVe} respectively: 
\bse\label{eq:skewlim}\begin{eqnarray}
&& (2p+{\hypotilde 0 u}-\wt{u})(1+\dot{u})+g(\wt{s}-\underaccent{\wtilde}{s})\dot{s}=2p\ ,  \label{eq:skewlima}\\ 
&& [(p+{\hypotilde 0 u})\wt{s}+(p-\wt{u})\underaccent{\wtilde}{s}]\dot{s}
=(\wt{s}-\underaccent{\wtilde}{s})(\dot{w}+s)\ , \label{eq:skewlimb}\\ 
&& [\wt{w}-{\hypotilde 0 w}-(p-u){\hypotilde 0 s} -(p+u)\wt{s}]\dot{s}
=({\hypotilde 0 s}-\wt{s})s(1+\dot{u})\ , \label{eq:skewlimc}\\
&& \left(p+u-\frac{\wt{w}}{\wt{s}}\right)\left(p-\wt{u}+
\frac{w}{s}\right)=p^2-\left(\frac{1}{s\wt{s}}+3e+gs\wt{s}\right)\ ,  \label{eq:skewlimd}\\ 
&& \left( 1+\frac{\dot{w}}{s}-\frac{w\dot{s}}{s^2}\right) 
\left(p-u+ \frac{{\hypotilde 0 w}}{{\hypotilde 0 s}}\right) + 
\left( p+{\hypotilde 0 u}-\frac{w}{s}\right) (1+\dot{u})% = \nn \\ 
%&& ~~~~~~~~~~~~ 
= 2p-\frac{\dot{s}}{s^2{\hypotilde 0 s}}+g{\hypotilde 0 s}\dot{s}\ ,  \label{eq:skewlime} 
\end{eqnarray}\ese 
cf. \cite{NijPutt}. The consistency of this system, where the undertilde 
\,$\ut{\cdot}$\, denotes the backward shift to the 
tilde-shift \,$\wt{\cdot}$\,, was discussed there as well: 
by direct computation, \eqref{eq:skewlime} is a direct consequence of 
\eqref{eq:skewlima}-\eqref{eq:skewlimd}, and hence is redundant.  Furthermore, it is easily checked, by taking 
the $t$-derivative of \eqref{eq:skewlimd} and back-substituting $\dot{u}$, $\dot{s}$ and $\dot{w})$ 
(which can be solved from the first three relations), 
that \eqref{eq:skewlimd} is compatible with those expressions. 

\subsection{Elliptic dressing chain}
Taking a straight continuum limit (in the sense of \cite{HJN}, Ch. 5) $q\to\infty$, 
$\wh{u}=u+\frac{1}{q}u'+\mathcal{O}(1/q^2)$, we obtain the differential-difference system\footnote{Note that \eqref{eq:EllDDKdVb} and \eqref{eq:EllDDKdVc} yield one and the same relation \eqref{eq:EllBTc} in this limit.}
\bse\label{eq:EllBT}\begin{align}
& (p+u-\wt{u})^2+u_x+\wt{u}_x= p^2+g(\wt{s}-s)^2\  , \label{eq:EllBTa} \\ 
%& \left(p+u-\frac{\wt{w}}{\wt{s}}\right) \wt{s}^2-\left( p-\wt{u}+\frac{w}{s}\right)s^2=
%(s\wt{s})'+2\left(u-\frac{w}{s}\right)s\wt{s}\  , \label{eq:EllBTb}\\ 
& \left(p+u-\frac{\wt{w}}{\wt{s}}\right) \wt{s}^2-\left( p-\wt{u}+\frac{w}{s}\right)s^2= 
(s\wt{s})_x+\left(u+\wt{u}-\frac{w}{s}-\frac{\wt{w}}{\wt{s}}\right)s\wt{s}\ , \label{eq:EllBTc}\\ 
& \left(p+u-\frac{\wt{w}}{\wt{s}}\right)\left(p-\wt{u}+
\frac{w}{s}\right)=p^2-\left(\frac{1}{s\wt{s}}+3e+gs\wt{s}\right)\ ,  \label{eq:EllBTd}\\ 
& \left(u+\frac{w}{s}\right)_x+\left( u-\frac{w}{s}\right)^2= \frac{1}{s^2}+3e+gs^2\ , \quad \left(\wt{u}+\frac{\wt{w}}{\wt{s}}\right)_x+
\left( \wt{u}-\frac{\wt{w}}{\wt{s}}\right)^2= \frac{1}{\wt{s}^2}+
3e+g\wt{s}^2\ . \quad 
\label{eq:EllBTe}
\end{align}\ese 
This is essentially also the B\"acklund transformation $(u,s,w)\to(\wt{u},\wt{s},\wt{w})$ 
for a corresponding elliptic continuous system with B\"acklund parameter $p$.  

{\bf Example:} Taking as seed solution $u=0,~w=1,~s=0$, the BT reduces to the set of 
equations:
\[ (p-\wt{u})^2+\wt{u}_x=p^2+g\wt{s}^2\ , \quad \wt{s}_x=2(p-\wt{u})\wt{s}+(p^2-3e)\wt{s}^2\ , 
\quad \wt{w}=p\wt{s}+1\  ,   \]
which comprises a coupled set of Riccati eqs. for $\wt{u}$ and $\wt{s}$.  This has a solution of the 
form
\[ \wt{u}=\frac{Ce^{2px}}{1+C(1-g/P^2)e^{2px}/(2p)}\ , \quad \wt{s}=\frac{1}{P}\wt{u}\ , \quad 
\wt{w}=1+p\wt{s}\ , \]
where $(p,P)$ are on the elliptic curve: $p^2=P+3e+g/P$.  The integration 
constant $C$ can contain additional variables, so if $C$ depends on the 
discrete variables $n'$ and $m'$ as 
\[ C(n,m)=\left(\frac{p'+p}{p'-p}\right)^{n'}
\left(\frac{q'+p}{q'-p}\right)^{m'} C(0,0)\  ,  \]
this 1-soliton solution also solves the elliptic lattice system 
\eqref{eq:EllDDKdV} with variables $n',m'$ and associated lattice parameters $p'$ and $q'$ respectively\footnote{Note that here we must 
distinguish between the role $p'$ and $q'$ play as lattice parameters 
and the role $p$ plays as B\"acklund parameter. }.

\section{Construction of the elliptic KdV system} 
\setcounter{equation}{0} 

We reiterate here briefly the construction of the Ell$\Delta\Delta$KdV 
system using an infinite `elliptic matrix' algebra. 

\subsection{Elliptic matrices} 

The construction of Ell$\Delta\Delta$KdV employs a realisation  
of an algebra $\mathcal A={\rm Mat}^0_{\Bbb Z}(\Bbb C)$ of 
centred infinite ``elliptic'' matrices. 
They are defined as quasi-graded algebra generated by two commuting 
index-raising operators $\Ld,\bL$ (and their adjoints $\tLd,\btL$)  
connected through the relation 
\begin{align*}\label{eq:PLrel}
\Ld^2=\bL+3e\boldsymbol{1}+g\bL^{-1}\ , \quad 
\tLd^2=\btL+3e\boldsymbol{1}+g\btL^{-1}\ ,
\end{align*}   
where $e,g\in\Bbb C$ are moduli of the elliptic curve given earlier. 
The elliptic matrices are defined through expressions of the form 
\begin{align*} 
 \bU=\sum_{i,j\in\mathbb{Z};\nu,\mu\in\{ 0,1\}} U_{2i+\nu,2j+\mu} 
 \tLd^\nu\,\btL^i\,\bO\,\Ld^\mu\,\bL^j\  , 
 \end{align*} 
where $\bO=\bee\tee$ is a rank 1 projector, and 
 $\tee\,\Ld^\mu\bL^j\tLd^\nu\btL^i\bee=\delta_{i,j}\delta_{\nu,\mu}$.   

The labeling of the elliptic matrix entries are defined as: 
\begin{align*} 
& U_{2i,2j}=\tee\,\bL^i\,\bA\,\btL^j\,\bee\ , \qquad  \quad 
U_{2i+1,2j}=\tee\,\bL^i\,\Ld\,\bA\,\btL^j\,\bee\ ,  \\ 
& U_{2i,2j+1}=\tee\,\bL^i\,\bA\,\tLd\,\btL^j\,\bee\ , \quad  
U_{2i+1,2j+1}=\tee\,\bL^i\,\Ld\,\bA\,\tLd\,\btL^j\,\bee\ .   
\end{align*} 
The following product rule applies to the algebra of elliptic 
matrices: 
\[ (\bU\cdot\bV)_{2i+\nu,2j+\mu}=\sum_{k\in\Bbb Z;\lambda\in\{ 0,1\} } U_{2i+\nu,2k+\lambda}
V_{2k+\lambda,2j+\mu}\ .   \] 
For later reference, note that we can also represent: 
\[ (a\boldsymbol{1}+\Ld)^{-1}= \frac{a\boldsymbol{1}-\Ld}{A-g/A} 
\left[ A(A\boldsymbol{1}-\bL)^{-1}-\frac{g}{A}(\frac{g}{A}\boldsymbol{1}-\bL)^{-1}\right] \ ,   
\] 
where $(a,A)$ is a point on the elliptic curve $\Gamma$: $a^2=A+3e+g/A$. 

\subsection{Linear dynamics} 

We have three types of linear dynamics acting on specific elements 
$\bC\in\mathcal A$, where the latter is a function of discrete and continuous 
variables: 
\begin{itemize}
\item{Discrete dynamics in discrete variables $n_\nu$ and lattice 
parameters $p_\nu$:} 
\bse\label{eq:Cdyns}
\be T_\nu\bC\,\left(p_\nu-\tLd\right)=(p_\nu+\bLd)\,\bC\   ,  
\label{eq:Cdynsa}
\ee 
\item{Discrete dynamics in discrete variables $N_\nu$ and lattice 
parameters $P_\nu$:} 
\be \mathcal{T}_\nu\bC\,\left(P_\nu-\btL\right) = (P_\nu-\bL)\,\bC\   , 
\label{eq:Cdynsb}\ee  
\item{Continuous dynamics in the parameters $p_\nu$:}   
\be \frac{\pl}{\pl p_\nu}\bC = n_\nu\left[ (p_\nu+\bLd)^{-1}\,\bC-
\bC\,\left(p_\nu-\tLd\right)^{-1}\right]\ ,  \label{eq:Cdynsc}\ee  
\item{Continuous dynamics in the parameters $p_\nu$:} 
\be \frac{\pl}{\pl P_\nu}\bC = N_\nu\left[ (P_\nu-\bL)^{-1}\,\bC-
\bC\,\left(P_\nu-\btL\right)^{-1}\right]\  .\label{eq:Cdynsd} \ee\ese  
\end{itemize} 

{\bf Remark:} The latter two are relevant in the elliptic KP case, cf. 
\cite{JennNij2014}. 
However, in the elliptic KdV case we impose the condition 
\[ \,^{t\!}\bC=\bC\ \Leftrightarrow   \bL\,\bC=\bC\,\btL\quad {\rm and} 
\quad \bLd\,\bC=\bC\,\tLd  \] 
in which case the latter dynamics (with parameters $P_\nu$) trivialises.  
In KP case, imposing the curve relation on the parameters: 
\[ p_\nu^2=P_\nu+3e+\frac{g}{P_\nu}\, \] 
will lead to further relations between $T_\nu$ shifts and $\mathcal{T}_\nu$ shifts.

\subsection{Elliptic Cauchy kernel} 

Introducing a formal elliptic Cauchy kernel $\bOm$ in the algebra of elliptic 
matrices , which is defined through the relations  
\begin{align*} 
\bOm\,\bLd+\tLd\,\bOm &= \bO-g\btL^{-1}\,\bO\,\bL^{-1}\    , \\
\bOm\,\bL-\btL\,\bOm &= \bO\,\bLd-\tLd\,\bO\   . 
%\Phi_\eta(\tLd)\cdot\bO\cdot\Phi_{\eta+\gm}(\Ld)\   . 
%\tII\cdot {\bf \Omega}+{\bf \Omega}\cdot {\bf I}+{\bf \Omega}= 0\  . 
\end{align*} 
These relations make sense in terms of the {\it symbols} of the 
operators\footnote{Note the distinction between the Cauchy kernels 
\[ \Omega(\kp,\ld)=\frac{1-g/(KL)}{k+l}=\frac{k-l}{K-L}  \quad \leftrightarrow 
\quad \mho(\kp,\ld)=\frac{k-l}{1-g/(KL)}=\frac{K-L}{k+l} \]
of the elliptic KdV system and those of the Landau-Lifschitz class of models that 
were used in \cite{DateLL,FuNij2022}, where the former is symmetric while the $\mho$ 
kernel is antisymmetric, leading to a Pfaffian structure in the solutions. 
}:   
\[
\bOm~~\leftrightarrow~~ \Omega(\kp_1,\kp_2)=\frac{k_1-k_2}{K_1-K_2}
=\frac{1-g/(K_1K_2)}{k_1+k_2} 
\]
in which we identify 
\[
k_i=\frac{1}{2}\frac{\wp'(\kp_i)}{\wp(\kp_i)-e}\quad ,\quad K_i=\wp(\kp_i)-e\ 
\quad (i=1,2)\ , 
\]
with $\wp(\kp)=\wp(\kp|2\oa,2\oa')$ the Weierstrass $\wp$-function with periods 
$2\oa$,$2\oa'$, and where $e=\wp(\oa)$. Note that 
$g=(e-e')(e-e'')$ (where $e,e',e''$ are the branch points).   

The index-raising operators can be interpreted in terms of their symbols 
as follows: 
\[ \Ld ~~~ \leftrightarrow ~~~ \zeta(\kp+\oa)-\zeta(\kp)-\zeta(\oa)=
\frac{1}{2}\frac{\wp'(\kp)}{\wp(\kp)-e}    \] 
\[ \tLd ~~~ \leftrightarrow ~~~ \zeta(\ld+\oa)-\zeta(\ld)-\zeta(\oa)=    
\frac{1}{2}\frac{\wp'(\ld)}{\wp(\ld)-e}    \] 
(where $\kp$ and $\ld$ can be thought of as the symbols for the matrices 
$\Lambda$ resp. $\,^{t\!}\Lambda$ corresponding to the rational case), and 
\[ \bL\quad\leftrightarrow\quad\wp(\kp)-e\  , \qquad  \btL\quad\leftrightarrow
\quad\wp(\ld)-e\ ,  \] 
where $\zeta(\cdot)$ is the Weierstrass $\zeta$-functions.

\subsection{Nonlinear structure} 

The nonlinear dynamics is given in terms of an elliptic matrix 
$\bU\in\mathcal A$ of the form  
\be\label{eq:Udef} 
\bU \equiv\bC\,\left(\boldsymbol{1}+\bOm\,\bC\right)^{-1}\quad   
\Leftrightarrow\quad  \bU=\bC-\bU\,\bOm\,\bC\  ,   
\ee  
in terms of which we have the following fundamental relations
\bse \be\label{eq:univcont} 
\partial\bU= \left(\boldsymbol{1}-\bU\,\bOm\right)\partial\bC\,\left(\boldsymbol{1}-\bOm\,\bU\right) \ee  
for any derivative $\partial$ (obeying the Leibniz rule) for which $\partial\bOm=0$, and 
\be\label{eq:univdisc}  
T\bU-\bU= \left(\boldsymbol{1}-(T\bU)\,\bOm\right)(T\bC-\bC)\,\left(\boldsymbol{1}-\bOm\,\bU\right)\ ,  \ee\ese 
for any shift operator for which $T\bOm=\bOm$. 
This leads to dynamical relations: 
\begin{itemize}
\item{Discrete dynamics in discrete variables $n_\nu$ and lattice 
parameters $p_\nu$:} 
\bse \label{eq:Udyns} 
\be\label{eq:Udynsdisc1} 
T_\nu\bU\,\left(p_\nu-\tLd\right)=(p+\bLd)\,\bU - 
(T_\nu\bU)\left( \bO-g\btL^{-1}\,\bO\,\bL^{-1}\right)\,\bU\   ,  \ee  
\item{Discrete dynamics in discrete variables $N_\nu$ and lattice 
parameters $P_\nu$:} 
\be\label{eq:Udynsdisc2}  
\mathcal{T}_\nu\bU\,\left(P_\nu-\btL\right) = (P_\nu-\bL)\,\bU 
+(\mathcal{T}_\nu\bU)\,(\bO\,\bLd-\tLd\,\bO)\,\bU\   , \ee  
\item{Continuous dynamics in the parameters $p_\nu$:}   
\begin{align}\label{eq:Udynscont1}
&\frac{\pl}{\pl p_\nu}\bU = n_\nu\left[ (p_\nu+\bLd)^{-1}\,\bU-
\bU\,\left(p_\nu-\tLd\right)^{-1}\right. \nn  \\ 
& \left. \qquad + \bU\,\left(p_\nu-\tLd\right)^{-1}
\left( \bO-g\btL^{-1}\,\bO\,\bL^{-1}\right)(p_\nu+\bLd)^{-1}\,\bU\, 
\right] \end{align} 
\item{Continuous dynamics in the parameters $P_\nu$:} 
\begin{align} \label{eq:Udynscont2}
&  \frac{\pl}{\pl P_\nu}\bU = N_\nu\left[ (P_\nu-\bL)^{-1}\,\bU-
\bU\,\left(P_\nu-\btL\right)^{-1}\right. \nn  \\ 
& \left. \qquad - \bU\,\left(P_\nu-\btL\right)^{-1}\,(\bO\,\bLd-\tLd\,\bO)\,
(P_\nu-\bL)^{-1}\,\bU \right]\  . 
\end{align}  \ese 
\end{itemize}

\subsection{KdV restriction and basic quantities} 

The above relations hold for the general KP case, cf. \cite{JennNij2014}. To restrict to 
the KdV case we have in addition the relation $\,^{t\!}\bU=\bU$~ 
which implies the algebraic relations 
\bse \label{eq:Ualg}\begin{align} 
&\bU\,\btL=\bL\,\bU-\bU\,\left( \bO\,\Ld-\tLd\,\bO\right)\,\bU~~~~~~  
\Leftrightarrow \label{eq:Ualga}\\     
&\bU\,\btL^{-1}=\bL^{-1}\,\bU+\bU\,\btL^{-1}\,\left(\bO\,\Ld-\tLd\,\bO\right)
\cdot\bL^{-1}\cdot\bU\   , \label{eq:Ualgb}
\end{align} \ese 
which are obtained by setting $P_\nu=0$ and $\mathcal{T}_\nu\bU=\bU$ 
in \eqref{eq:Udynsdisc2}. 
The latter two equations also guarantee that the system is invariant under 
the symmetry 
\[ p_\nu \leftrightarrow -p_\nu\ , \quad T_\nu\bU \leftrightarrow T_\nu^{-1}\bU\ . \] 
The basic dependent variables of the elliptic system are given by: 
\bse\label{eq:usvwh}\begin{eqnarray} 
&&u\equiv\tee\,\bU\,\bee\ ,\quad s\equiv\tee\,\bL^{-1}\bU\,\bee\ ,  
\quad h\equiv\tee\,\bL^{-1}\,\bU\,\tbL^{-1}\bee\ ,   \\ 
&&v\equiv 1-\tee\,\bL^{-1}\,\bLd\,\bU\,\bee\ , \quad 
w\equiv 1+\tee\,\bL^{-1}\bU\,\tLd\,\bee\ .  
\end{eqnarray} \ese 
Fixing $p_1=p$, $p_2=q$, and denoting 
$T_1f=\wt{f}$, $T_2f=\wh{f}$, we get the basic relations 
that yield the elliptic $\Delta\Delta$KdV system are: 
\bse\label{eq:Miura}\begin{align}  
& \left(p-gh+\frac{\wt{v}}{\wt{s}}\right)\wt{s}=\left(p-\wt{u}+\frac{w}{s}\right)s\ , \quad 
\left( p+g\wt{h}-\frac{v}{s}\right)s=\left(p+u-\frac{\wt{w}}{\wt{s}}\right)\wt{s}\ , \label{eq:Miuraa} \\ 
& U_{1,-1}=(1-vw)/s\ , \quad 
\wt{U}_{-1,-2}+U_{-1,-2}= p(\wt{h}-h)-g\wt{h}h+\wt{s}s\ , \label{eq:Miurab} \\ 
& p(v-\wt{v})=gvh+gs\left( U_{-2,-1}+\wt{U}_{-2,-1}\right)+\wt{U}_{-1,1}+3es
+\wt{v}u  \nn  \\ 
&=g\wt{v}\wt{h}+g\wt{s}\left( U_{-2,-1}+\wt{U}_{-2,-1}\right)+U_{-1,1}+
3e\wt{s} +v\wt{u}\ , \label{eq:Miurac} \\
%\qquad , \qquad  \bblu U_{2i,2j+1}=\tee\,\bL^i\,\bU\,\tLd\,\tbL^j\,\bee\ecl\ ,  \\ 
& p(\wt{w}-w)=g\wt{h}w-\wt{s}\left( U_{0,1}+\wt{U}_{0,1}\right)
+U_{-1,1}+3e\wt{s}+ \wt{w}\wt{u}  \nn  \\ 
&=gh\wt{w}-s\left( U_{0,1}+\wt{U}_{0,1}\right)+\wt{U}_{-1,1}+3es
+wu\ , \label{eq:Miurad} \\  
& U_{0,1}+\wt{U}_{0,1}= p(\wt{u}-u)+\wt{u}u-g\wt{s}s\ , \label{eq:Miurae}
\end{align} \ese 
where $U_{0,1}=\tee\,\bU\,\tLd\,\bee$, $U_{-1,1}=\tee\,\bL^{-1}\,\Ld\,\bU\,\tLd\,\bee$ and  $U_{-2,1}=\tee\,\bL^{-1}\,\bU\,\tLd\,\bee$.

\subsection{Construction of Lax pairs (EllKdV case)} 
Finally, in order to obtain the relevant Lax pairs, we introduce also the 
infinite vectors $\buk$ by 
\begin{equation} \label{eq:uk}  
\buk = \left( \boldsymbol{1}- \bU\cdot\bOm\right)\,\bc_k\rho_k\  , 
\end{equation}  
in which $\rho_k$ (the ``plane-wave factor'') contains the dynamical 
variables:  
\be\label{eq:pwf} 
\rho_k=\left[ \prod_\nu \left( \frac{p_\nu+k}{p_\nu-k}\right)^{n_\nu} 
\right]\rho_k(\boldsymbol{0})\  .   \ee 
The $\bc_k$ are eigenvectors of the operators $\bL,\bLd$: 
\[ \bLd\,\bc_k=k\bc_k\ , \quad \bL\,\bc_k=K\bc_k\  . \] 
We can derive 
the basic set of (linear) equations for the vector $\buk$: 
\bse\label{eq:ukdyns}\begin{align}
(p_\nu-k)T_\nu\bu_k &= \left[ (p_\nu+\bLd)-(T_\nu\bU)\,\left(\bO-g\tbL^{-1}\,
\bO\,\bL^{-1}\right)\right]\,\buk \    ,  \\ 
(p_\nu+k)\bf u_k &= \left[ (p_\nu-\bLd)+\bU\,\left(\bO-g\tbL^{-1}\,\bO 
\,\bL^{-1}\right)\right]\,T_\nu\buk \   , \label{eq:bukrelb}  
\end{align}\ese 
as well as the linear ``algebraic'' relations 
\bse\label{eq:ukalg}\begin{align}
&K\,\buk =\left[\bL- \bU\cdot\left( \bO\cdot\Ld-\tLd\cdot\bO\right)\right]\cdot\buk~~~~~~  
\Leftrightarrow \\     
&\frac{1}{K}\,\buk=\left[ \bL^{-1}+\bU\cdot\tbL^{-1}\cdot\left(\bO\cdot\Ld-\tLd\cdot\bO\right)
\cdot\bL^{-1}\right]\cdot\buk\   , 
\end{align} \ese 
where $K$ and $k$ are related through the elliptic curve $k^2=K+3e+g/K$~. 

We have also the linear differential relations: 
\begin{align}\label{eq:ukdiff}
\frac{\pl}{\pl p_\nu}\buk = n_\nu\left[\frac{1}{k-p_\nu}+ 
(p_\nu+\bLd)^{-1} 
+\bU\,(p_\nu-\tLd)^{-1}\,\left( \bO-g\tbL^{-1}\,\bO\bL^{-1}\right)
\,(p_\nu+\bLd)^{-1}\right]\,\buk\ .    
\end{align} 
\iffalse 
or alternatively the hierarchy of differential relations: 
\begin{equation}    
\frac{\pl}{\pl x_j}\buk = \Ld^j\cdot\buk-(-k)^j\buk-\bU\cdot 
\left( \bO_j-f\tbP^{-1}\cdot\bO_j\cdot\bP^{-1}\right)\cdot\buk\quad,\quad j\in\mathbb{Z}\  .   
\end{equation}    
\fi 
{}From these equations one can derive the relevant Lax pairs.

\subsection{$\tau$-function} 

The $\tau$-function, in 
the present context is identified with the infinite determinant: 
\be\label{eq:tau} 
\tau={\det}\!_{\mathcal A}\left( {\bf 1}+\bOm\cdot\bC\right) \  , 
\ee 
where the determinant ${\det}\!_{\mathcal A}$~ in the elliptic algebra 
can be given a proper meaning by means of the expansion of the 
determinant in trace-invariants of powers of the elliptic matrix 
$\bOm\bC$ noting that the Cauchy kernel $\bOm$ contains a 1-dimensional 
projector $\bO$. Assuming some natural 
properties we can derive (using a Weinstein-Aronszajn type formula):  
\begin{align} 
\frac{T_\nu\tau}{\tau}&={\det}\!_{\mathcal A}\left\{ {\bf 1}+
\left(\bO- g\tbL^{-1}\,\bO\,\bL^{-1}\right)\,\bU\,(p_\nu-\tLd)^{-1}\right\}\nn\\ 
&=\left|\begin{array}{ccc}
1+\tee\,\bU\,(p_\nu-\tLd)^{-1}\bee &\quad& 
\tee\,\bL^{-1}\,\bU\,(p_\nu-\tLd)^{-1}\bee\\
-g\tee\,\bU\,\tbL^{-1}\,(p_\nu-\tLd)^{-1}\bee &\quad&
1-g\tee\,\bL^{-1}\,\bU\,\tbL^{-1}\,(p_\nu-\tLd)^{-1}\bee 
\end{array}\right|\  . \label{eq:WeinAronsz} 
\end{align} 
The basic differential equation for the $\tau$-function reads: 
\begin{align}\label{eq:taudiffrel}
\frac{\pl}{\pl p_\nu}\ln\tau & =-n_\nu\left[ \tee\,(p_\nu+\bLd)^{-1}\,
\bU\,(p_\nu-\tLd)^{-1}\bee \right. \nn \\ 
&\qquad\qquad \left. -g \tee\,(p_\nu+\bLd)^{-1}\,\bL^{-1}\, 
\bU\,\tbL^{-1}\,(p_\nu-\tLd)^{-1}\bee\right]\  .  
\end{align}
These relation invite the introduction of a number of parameter-dependent 
quantities (with arbitrary parameter $a$), namely 
\bse\label{eq:defs}\begin{eqnarray} 
v_a\equiv 1-\tee\,(a+\bLd)^{-1}\,\bU\,\bee~~~~&,&~~~~ 
s_a\equiv\tee\,(a+\bLd)^{-1}\,\bL^{-1}\,\bU\,\bee~~,  \\ 
w_a\equiv\tee\,(a+\bLd)^{-1}\,\bU\,\tbL^{-1}\bee ~~~~&,&~~~~ 
h_a\equiv 1+g\tee\,(a+\bLd)^{-1}\,\bL^{-1}\,\bU\,\tbL^{-1}\bee~~,    
\end{eqnarray}\ese 
as well as 2-index variables (with parameters $a,b$), namely
\bse\label{eq:defs2}\begin{align} 
&s_{a,b}\equiv \tee\,(a+\bLd)^{-1}\,\bU\,(b+\tLd)^{-1}\bee ~~,~~   
t_{a,b}\equiv\tee\,(a+\bLd)^{-1}\,\bL^{-1}\,\bU\,(b+\tLd)^{-1}\bee
\ ,   \\ 
&u_{a,b}\equiv g\tee\,(a+\Ld)^{-1}\,\bL^{-1}\cdot\bU\,\tbL^{-1}
\cdot(b+\tLd)^{-1}\bee\ .     
\end{align}\ese 
To derive closed-form relations for the $\tau$-function \eqref{eq:tau} we need to explore further the dynamical relations between these qantities. 
This is what we will do in the next section. 

\section{Matrix lattice systems} 

In this section we explore further the parameter-dependent quantities 
\eqref{eq:defs} and \eqref{eq:defs2} and derive related matricial 
equations. 

\subsection{Auxiliary variables} 

Apart from the principal new objects \eqref{eq:defs} and \eqref{eq:defs2} 
we also need the following auxiliary variables:
\bse\label{eq:pqrt} 
\begin{align}\label{eq:rq} 
& r_a\equiv a-\tee\,(a+\Ld)^{-1}\cdot\bU\cdot\tLd\,\bee\  ,  
\qquad q_a\equiv \tee\, 
(a+\Ld)^{-1}\cdot\bL^{-1}\cdot\bU\cdot\tLd\,\bee\   ,  \\ 
& p_a\equiv\tee\,(a+\Ld)^{-1}\cdot\bU\cdot\tLd\cdot\btL^{-1}\bee\ , 
\quad t_a\equiv a+g\tee\,(a+\Ld)^{-1}\cdot\bL^{-1}\cdot\bU\cdot\tLd\cdot\btL^{-1}\bee\ . 
\label{eq:pt} 
\end{align} \ese 
These variables can be assembled in the 2$\times$2 matrices:
\begin{equation}\label{eq:VW}  
\bV_a\equiv\left(\begin{array}{cc} 
v_a & -w_a \\ gs_a & h_a \end{array}\right) \qquad  ,\qquad  
\bR_a\equiv\left(\begin{array}{cc} 
r_a & -p_a \\ gq_a & t_a \end{array}\right) \  .  
\end{equation} 
The latter variables can be expressed (see Appendix D) in terms of the 
variables \eqref{eq:defs} as follows: 
\bse \label{eq:pqrtrels}\begin{eqnarray} 
&&p_a=\frac{s_a-vw_a}{s}\  ,  \qquad q_a=\frac{ws_a-w_a}{s}\  ,  \label{eq:rqb}\\  
&&t_a=\frac{v_a-vh_a+(a^2-3e)s_a}{s}\  ,  \qquad 
r_a=\frac{wv_a-h_a+(a^2-3e)w_a}{s}\   .  \label{eq:rqc} 
\end{eqnarray}\ese  
Note that the relations \eqref{eq:pqrtrels} can be inverted: 
\bse \label{eq:vwshrels}\begin{eqnarray} 
&&w_a=s\frac{q_a-wp_a}{vw-1}\  ,  \qquad s_a=s\frac{vq_a-p_a}{vw-1}\  ,  \label{eq:rqb}\\  
&&h_a=s\frac{r_a-wt_a+(a^2-3e)q_a}{vw-1}\  ,  \qquad 
v_a=s\frac{vr_a-t_a+(a^2-3e)p_a}{vw-1}\   .  \label{eq:rqc} 
\end{eqnarray}\ese  
Writing \eqref{eq:pqrtrels} in matrix form, we have 
\begin{align}\label{eq:Rrel}   
& \boldsymbol{R}_a\left(\begin{array}{cc}0&1\\ g&0 \end{array}\right)= \nn \\ 
& = \frac{1}{s}\left[ \boldsymbol{V}_a \left(\begin{array}{cc} 
\tfrac{1}{2}(a^2-3e) & w \\ -gv & -\tfrac{1}{2}(a^2-3e)\end{array}\right) 
- \left(\begin{array}{cc} 
\tfrac{1}{2}(a^2-3e) & 1 \\ -g & -\tfrac{1}{2}(a^2-3e)\end{array}\right)\boldsymbol{V}_a
\right]\ ,  
\end{align}
expressing $\bR_a$ in terms of $\bV_a$. 

The 2-index variables can be collected in the 2$\times$2 matrix: 
\be\label{eq:Z}   
%\boldsymbol{V}_a=\left( \begin{array}{cc} 
%v_a & -w_a\\ gs_a & h_a \end{array}\right) \ , \quad 
\boldsymbol{Z}_{a,b} = \left( \begin{array}{cc} 
s_{a,b} & t_{b,a} \\ -g t_{a,b} & -u_{a,b}\end{array} \right)\ ,  
\ee 
where $s_{a,b}=s_{b,a}$, $u_{a,b}=u_{b,a}$ (while $t_{a,b}\neq t_{b,a}$!), 
and which obeys the symmetry 
~$G\boldsymbol{Z}_{b,a}^T G^{-1}=\boldsymbol{Z}_{a,b}$~, where 
$G={\rm diag}(1,-g)$.  
Due to \eqref{eq:Ualgb}, the entries of the matrix $\boldsymbol{Z}_{a,b}$ are subject to the relations:
\bse\label{eq:Zconds}\bea 
&& s_{a,b}-(a^2-3e)t_{a,b}+u_{a,b}=h_a p_b-t_a w_b\ , \\ 
&& s_{a,b}-(a^2-3e)t_{b,a}+u_{a,b}=v_a q_b-r_a s_b\ . 
\eea\ese 
Whereas $s_{a,b}$ and $u_{a,b}$ are symmetric in the indices, 
$t_{a,b}$ and $t_{b,a}$ differ by\footnote{The conditions \eqref{eq:Zconds} can be symmetrised by setting 
\[
s_{a,b}-(a^2-3e)\ell_{a,b}+u_{a,b}=\frac{h_a s_b-v_a w_b}{s}\  ,  \quad 
{\rm where} \quad 
\ell_{a,b}\equiv t_{a,b}-\frac{s_a w_b}{s}=\ell_{b,a}\  . 
\] }  
\bse\label{eq:Zconstrs}\begin{equation}\label{eq:tcond} 
t_{a,b}-t_{b,a}= \frac{1}{s}(s_a w_b-w_a s_b)\  , \end{equation} 
using \eqref{eq:pqrtrels}, which together with \eqref{eq:Zconds}, which can be 
rewritten as 
\begin{equation}\label{eq:Zcond}
s_{a,b}-(a^2-3e)t_{a,b}+u_{a,b}=\frac{1}{s}\left[ h_a s_b-v_a w_b-(a^2-3e)s_a w_b\right]\ ,   
% & t_{a,b}-t_{b,a}= \frac{1}{s}(s_a w_b-w_a s_b)\  , 
\end{equation}\ese 
form a set the constraints on the entries of the matrix $\bZ_{a,b}$. 

Finally, we can present a ``strange'' identity:
\bse\label{eq:strange}\begin{equation} 
a-b+(a^2-b^2) u_{a,b}+g(t_{a,b}-t_{b,a}) =  t_a h_b-t_b h_a   
\end{equation} 
which is obtained from \eqref{eq:Ualgb} by sandwiching the elliptic matrix 
relations a factor $(a+\Ld)^{-1}\bL^{-1}$ from the left and a factor 
$(b+\tLd)^{-1}\btL^{-1}$ from the left and using the curve relation.  
It can be recast in the form 
\begin{equation} 
1+(a+b)u_{a,b} =\frac{v_a h_b-v_b h_a+(a^2-3e)s_a h_b-(b^2-3e)s_b h_a+g(w_a s_b-w_b s_a)}{(a-b) s} 
\end{equation}\ese  
for $a\neq b$, which tells us that $u_{a,b}$ can be expressed entirely in terms 
of the entries of the matrices $\bV_a$ and $\bV_b$. The same holds true for the entry 
$t_{a,b}$, which, from \eqref{eq:Zconds} using \eqref{eq:pqrtrels}, can be expressed  
as
\begin{align}
t_{a,b} &= \frac{1}{s}s_a w_b +\frac{1}{s}\,\frac{(h_a s_b-s_a h_b)+(w_a v_b-v_a w_b)}{b^2-a^2} \ . 
%\nn \\ 
% &= \frac{1}{s}s_a w_b + \frac{1}{(a^2-b^2)s}\, {\rm tr} \left\{\left(\begin{array}{cc} 0 & g^{-1} \\ 
%1 & 0 \end{array}\right) \bV_a\,\cof(\bV_b)\right\}\ .  
\end{align}
For the remaining entry $s_{a,b}$ we have, by back-substituting the above expressions 
into either one of \eqref{eq:Zconds},  the following result: 
\begin{align}
1-(a+b)s_{a,b} =\frac{v_a h_b-v_b h_a+(a^2-3e)w_a v_b-(b^2-3e)w_b v_a+g(w_a s_b-w_b s_a)}{(a-b) s}\ . 
\end{align} 
Thus, all entries of the matrix $\bZ_{a,b}$ (when $a\neq b$) can be expressed in terms of the entries of the matrices $\bV_a$ and $\bV_b$ and $s$. They can be collected in the following matrix form 
\bse\label{eq:matformZ}\be
\boldsymbol{1}-(a+b)\bZ_{a,b} = \frac{1}{(a-b)s}({\rm tr}\otimes{\rm id})\left\{ ( \bV_a\,\cof(\bV_b)\otimes\boldsymbol{1})\mathbb{R}_{a,b}
-(a^2-b^2)(\bV_a\,\bS_-\cof(\bV_b)\otimes\boldsymbol{1})\mathbb{S}\right\}   
\ee 
where 
\bea\label{eq:RSmat}
\mathbb{R}_{a,b}&=& \bsg_3\otimes \boldsymbol{1}+ i\bsg_2\,G^{-1}\otimes 
G\bsg_1 + (a^2-b^2) \bS_-\otimes \bF \nn \\ 
&& \qquad +(a^2-3e) \frac{1}{g}\bE\otimes\bS_- -(b^2-3e) \bF\otimes\bS_+\  ,  \\ 
\mathbb{S}&=&  \frac{1}{g}\bS_+\otimes \bE 
+\bS_-\otimes \bF+\frac{1}{g}\bE\otimes\bS_- +\bF\otimes\bS_+\ .  
\eea\ese  
Here, again $G={\rm diag}(1,-g)$, and 
\[ \bE=\left(\begin{array}{cc} 0 & 1 \\ 0 & 0\end{array} \right)\ , 
\quad 
\bF=\left(\begin{array}{cc} 0 & 0 \\ 1 & 0\end{array} \right)\ , \quad 
\bS_+=\left(\begin{array}{cc} 1 & 0 \\ 0 & 0\end{array} \right)\ , \quad 
\bS_-=\left(\begin{array}{cc} 0 & 0 \\ 0 & 1\end{array} \right)\ , 
\]
the $\boldsymbol{\sigma}_i$ ($i=1,2,3$) are the 
standard Pauli matrices and ~$\cof(\bV)=\det(\bV)\,\bV^{-1}$~ denotes the matrix of cofactors of a 2$\times$2 matrix $\bV$.  
\iffalse 
An alternative to eq. \eqref{eq:matformZ} is the following: 
\bse\label{eq:altmatformZ}\be
\boldsymbol{1}-(a+b)\bZ_{a,b} = \frac{1}{(a-b)s}({\rm tr}\otimes{\rm id})\left\{  \Big(\bV_a\bS_+\cof(\bV_b)\otimes\boldsymbol{1}\Big)
\mathbb{R}^+_{a,b} + \Big(\bV_a\bS_-\cof(\bV_b)\otimes\boldsymbol{1}\Big)
\mathbb{R}^-_{a,b}\right\}   
\ee 
where 
\bea\label{eq:R+R-mat}
\mathbb{R}^+_{a,b}&=& \bsg_3\otimes \boldsymbol{1}+ i\bsg_2\,\bG^{-1}\otimes \bG\bsg_1 + (a^2-b^2) \bS_-\otimes \bF \nn \\ 
&& \qquad +(a^2-3e) \frac{1}{g}\bE\otimes\bS_- -(b^2-3e) \bF\otimes\bS_+\  ,  \\ 
\mathbb{R}^-_{a,b}&=& \bsg_3\otimes \boldsymbol{1}+ i\bsg_2\,\bG^{-1}\otimes \bG\bsg_1 - (a^2-b^2) \bS_+\otimes \frac{1}{g}\bE \nn \\ 
&& \qquad -(a^2-3e)\bF\otimes\bS_+ +(b^2-3e) \frac{1}{g}\bE\otimes\bS_- \  ,  
\eea\ese 
\fi 

\paragraph{Remark:} It is interesting to see from the definitions (\ref{eq:defs2}) that the matrix 
$\bZ_{a,b}$ can be expanded in inverse powers of $b$ as follows 
\[ \bZ_{a,b}=\frac{1}{b}\left( \boldsymbol{1}-\bV_a\right) 
-\frac{1}{b^2}\left( a\boldsymbol{1}-\bR_a\right) + \cdots  \] 
and also 
\[ \bV_{a}= \boldsymbol{1}- \frac{1}{a}\bS+\frac{1}{a^2}\left(\begin{array}{cc}
U_{1,0} & w-1 \\ g(v-1) & -g U_{-1,-2} \end{array}\right)+\cdots \  , \]
which shows how the various quantities we have introduced are related.

\subsection{Matrix equations}

Using the relations from the elliptic matrix structure, we can derive 
the matrix relations 
\bse \label{eq:RSmat} \begin{align} 
\wt{\boldsymbol{R}}_a &= \wt{\boldsymbol{V}}_a (p\boldsymbol{1}+\boldsymbol{S})
-(p-a)\boldsymbol{V}_a\ , \label{eq:RSmata} \\
\boldsymbol{R}_a &= \boldsymbol{V}_a (-p\boldsymbol{1}+\wt{\boldsymbol{S}})
+(p+a)\wt{\boldsymbol{V}}_a\ , \label{eq:RSmatb}
\end{align} \ese 
in which 
\begin{equation}\label{eq:Smat}
\bS\equiv\left(\begin{array}{cc} u&s\\ -gs&-gh \end{array}\right)\ , 
\quad {\rm which\ \ obeys}\quad \bS^T=G^{-1}\bS G\ , 
% {\rm where} \quad G={\rm diag}(1,-g)\ , 
\end{equation}
and similar relations involving the shift $\wh{\phantom{a}}$ and lattice parameter $q$. 
From the latter we can derive the \textit{matrix Miura type relations}: 
\bse\label{eq:SVrels}\begin{align} 
(p-q)\boldsymbol{1}+\wh{\boldsymbol{S}}-\wt{\boldsymbol{S}} &= 
\boldsymbol{V}_a^{-1}\left((p+a)\wt{\boldsymbol{V}}_a-(q+a)\wh{\boldsymbol{V}}_a\right) \label{eq:SVrelsa}  \\ 
&=  \wh{\wt{\boldsymbol{V}}}_a^{-1}\left((p-a)\wh{\boldsymbol{V}}_a-
(q-a)\wt{\boldsymbol{V}}_a\right)\ , \label{eq:SVrelsb} \\  
(p+q)\boldsymbol{1}+\boldsymbol{S}-\wh{\wt{\boldsymbol{S}}} &= 
\wt{\boldsymbol{V}}_a^{-1}\left((p-a)\boldsymbol{V}_a+(q+a)\wh{\wt{\boldsymbol{V}}}_a\right) \label{eq:SVrelsc} \\ 
&=  \wh{\boldsymbol{V}}_a^{-1}\left((p+a)\wh{\wt{\boldsymbol{V}}}_a+(q-a)\boldsymbol{V}_a\right) \label{eq:SVrelsd}
\end{align} \ese 
The equalities on the right-hand sides of \eqref{eq:SVrelsa} and \eqref{eq:SVrelsb}, or equivalently \eqref{eq:SVrelsc} and \eqref{eq:SVrelsd}, 
form a \textit{matrix lattice modified KdV system}\footnote{In fact, setting $g=0$ (when the elliptic curve degenerates) in \eqref{eq:VW} 
we get ~$\bV_a=\left(\begin{array}{cc} v_a & -w_a\\ 0 & 1\end{array}\right)$~ and then from \eqref{eq:SVrels} we obtain the 
lattice MKdV 
\[ \left[(p+a)\wt{v}_a-(q+a)\wh{v}_a\right]\wh{\wt{v}}_a= 
\left[(p-a)\wh{v}_a-(q-a)\wt{v}_a\right]v_a\ , 
\] 
together with the relations 
\[ \frac{\wh{w}_a}{\wt{w}_a}=\frac{(p+a)\wh{\wt{v}}_a+(q-a)v_a}{(q+a)\wh{\wt{v}}_a+(p-a)v_a}\  , \quad 
\frac{\wh{\wt{w}}_a}{w_a}=\frac{(p-a)\wh{v}_a-(q-a)\wt{v}_a}{(p+a)\wt{v}_a-(q+a)\wh{v}_a}\  , 
\]
which determines $w_a$.  Furthermore, when $g=0$ we have $\bZ_{a,b}= 
\left(\begin{array}{cc} s_{a,b}&t_{b,a} \\ 0 & 0 \end{array}\right)$, and 
\eqref{eq:matZeq} reduces to the NQC equation 
\[ \frac{1+(p-a)s_{a,b}-(p+b)\wt{s}_{a,b}}{1+(q-a)s_{a,b}-(q+b)\wh{s}_{a,b}}= \frac{1+(q-b)\wt{s}_{a,b}-(q+a)\wh{\wt{s}}_{a,b}}{1+(p-b)\wh{s}_{a,b}-(p+a)\wh{\wt{s}}_{a,b}}= 
\frac{(p-a)t_{b,a}-(p+q)\wt{t}_{b,a}+(q+a)\wh{\wt{t}}_{b,a}}{(q-a)t_{b,a}-(q+p)\wh{t}_{b,a}+(p+a)\wh{\wt{t}}_{b,a}}\ , 
\]
where the latter equality determines $t_{b,a}$. 
}. 
\iffalse 
but which are also subject to relations stemming from the identification  
\begin{align}\label{eq:Rrel}   
& \boldsymbol{R}_a\left(\begin{array}{cc}0&1\\ g&0 \end{array}\right)= \nn \\ 
& = \frac{1}{s}\left[ \boldsymbol{V}_a \left(\begin{array}{cc} 
\tfrac{1}{2}(a^2-3e) & w \\ -gv & -\tfrac{1}{2}(a^2-3e)\end{array}\right) 
- \left(\begin{array}{cc} 
\tfrac{1}{2}(a^2-3e) & 1 \\ -g & -\tfrac{1}{2}(a^2-3e)\end{array}\right)\boldsymbol{V}_a
\right]\ ,  
\end{align}
which involves also $v$~ and $w$. 
\fi 

Setting $a=p$ in \eqref{eq:SVrelsb} and \eqref{eq:SVrelsc} we derive for 
$\bS$ the \textit{matrix potential KdV equation}: 
\begin{equation} \label{eq:matpotKdV}
[(p-q)\boldsymbol{1}+\wh{\boldsymbol{S}}-\wt{\boldsymbol{S}}]\,
[(p+q)\boldsymbol{1}+\boldsymbol{S}-\wh{\wt{\boldsymbol{S}}}]= 
(p^2-q^2)\boldsymbol{1}\ , 
\end{equation} 
while the equalities on the r.h.s. of \eqref{eq:SVrels} lead to the 
matrix lattice mKdV equation. 
Note that these matrix equations are not free, but subject to constraints 
on the entries related to the elliptic curve. 
Furthermore, as a consequence of the relations (\ref{eq:RSmat}), we have also 
the relation:
\begin{equation}
2p\wt{\bV}_{\!p}=\bV_{\!p}\,\left( 2p\boldsymbol{1}+\ut{\bS}-\wt{\bS}\right)\  ,  
\end{equation}
which will play a role in the next section. 
Also, the relation for the $\tau$-function, (\ref{eq:tau}), 
can be expressed in terms of these matrices, namely 
\begin{equation}\label{eq:dmattau}  
\frac{\wt{\tau}}{\tau}=\det(\bV_{\!-p})=\det(\wt{\bV}_{\!p}^{-1})\  .  
\end{equation}

%Rewriting \eqref{eq:zgkrels} in terms of t
The 2-indexed matrices \eqref{eq:Z} obey  
the following dynamical relations can be derived: 
%\[ \det\left(\boldsymbol{V}_{-p}\right) = \frac{\wt{\tau}}{\tau}\ , \qquad 
%\det\left(\wt{\boldsymbol{V}}_p\right)=\frac{\tau}{\wt{\tau}}\ , \] 
%and 
\begin{equation}  \label{eq:ZVrel}
\boldsymbol{1}+(p-a)\boldsymbol{Z}_{a,b}-(p+b)\wt{\boldsymbol{Z}}_{a,b}=\wt{\boldsymbol{V}}_a G
\boldsymbol{V}_b^T G^{-1}\  , %\quad {\rm with}\quad 
%G=\left( \begin{array}{cc} 1 & 0 \\ 0 & -g\end{array}\right)\ , 
\end{equation}  
(and a similar relations with $\wt{\phantom{a}}$ replaced by $\wh{\phantom{a}}$ and 
$p$ replaced by $q$) with $G={\rm diag}(1,-g)$ as before. 
In particular,  the \textit{inversion 
relation} ~$\wt{\bV}_{\!p}\,G\,\bV_{\!-p}^T\,G^{-1}=\boldsymbol{1}$~ follows 
directly by setting $a=p$ and $b=-p$ in \eqref{eq:ZVrel}.  
%\begin{equation}\label{eq:inverse} 
%\wt{\bV}_{\!p}\,G\,\bV_{\!-p}^T\,G^{-1}=\boldsymbol{1}~~~,~~~~~
%G\equiv{\rm diag}(1,-g)~~~. 
%\end{equation}

Using the identity ~$G\boldsymbol{Z}_{b,a}^T G^{-1}=\boldsymbol{Z}_{a,b}$~  
the following \textit{matrix NQC equation} holds: % for $\mathbold{Z}_{a,b}$:  
\begin{align} \label{eq:matZeq} 
& [\boldsymbol{1}+(p-a)\boldsymbol{Z}_{a,b}-(p+b)\wt{\boldsymbol{Z}}_{a,b}]\,
[\boldsymbol{1}+(q-a)\boldsymbol{Z}_{a,b}-(q+b)\wh{\boldsymbol{Z}}_{a,b}]^{-1} 
\nn \\  
& \quad =[\boldsymbol{1}+(q-b)\wt{\boldsymbol{Z}}_{a,b}-(q+a)\wh{\wt{\boldsymbol{Z}}}_{a,b}]\,
[\boldsymbol{1}+(p-b)\wh{\boldsymbol{Z}}_{a,b}-(p+a)\wh{\wt{\boldsymbol{Z}}}_{a,b}]^{-1}\ , 
\end{align} 
subject to the constraint \eqref{eq:matformZ} on the form of the matrix $\boldsymbol{Z}_{a,b}$, containing the moduli of the elliptic curve $\Gamma$.

\subsection{Cauchy matrix (soliton type) solutions} 
In the context of the soliton solutions the $\tau$-function is given by {\small 
\begin{align} \label{eq:tauform}
\tau=\det\left(\boldsymbol{I}+\boldsymbol{M}\right)  
 =\sum_{m=0}^N\sum_{{J_m\subset \{1,\cdots,N\}}\atop |J_m|=m}  
(-1)^m \frac{\sigma(\omega+2\sum_{i\in J_m}\kappa_i)}{\sigma(\omega)\prod_{i\in J_m } \sigma(2\kappa_i)K_i } 
\left[ \prod_{i<j\in J_m}\left(\frac{\sigma(\kappa_i-\kappa_j)}{\sigma(\kappa_i+\kappa_j)} \right)^2 \right]\prod_{i\in J_m} r_ic_i\ . 
\end{align} } 
Here $\bM$ is the elliptic Cauchy matrix and $r_i$ are plane-wave factors, 
respectively  
\[ (\bM)_{i,j}= \frac{1-g/(K_iK_j)}{k_i+k_j}r_i c_j= \frac{k_i-k_j}{K_i-K_j}r_i c_j\ , 
\quad r_i(n,m)=\left(\frac{p+k_i}{p-k_i}\right)^n\left(\frac{q+k_i}{q-k_i}\right)^m\ , \]
with rapidities $(k_i,K_i)$, with uniformising variables $\kappa_i$, on the elliptic curve 
$k_i^2=K_i+3e+g/K_i$, and the $c_j$ are arbitrary constants. The $r_i$ can also be written as
\[ r_i(n,m)
=\left(\frac{\sigma(\kappa_i-\alpha+\omega)\,\sigma(\kappa_i+\alpha)}{\sigma(\kappa_i+\alpha+\omega)\,\sigma(\kappa_i-\alpha)} e^{2\eta\alpha}\right)^n
\left(\frac{\sigma(\kappa_i-\beta+\omega)\,\sigma(\kappa_i+\beta)}{\sigma(\kappa_i+\beta+\omega)\,\sigma(\kappa_i-\beta)} e^{2\eta\beta}\right)^m\ ,  \]
where $\omega$ is again the fixed half-period, $\eta=\zeta(\omega)$ and $\alpha$, $\beta$ 
are such that $p=\zeta(\alpha+\omega)-\zeta(\alpha)-\eta$, $q=\zeta(\beta+\omega)-\zeta(\beta)-\eta$. 
The terms in the expansion given by right-hand side of \eqref{eq:tauform} arises 
from the Frobenius formula for an elliptic Cauchy determinant, \cite{Frob}. 
The soliton solutions for the variables $u$, $v$, $w$, $s$ and $h$ were 
already presented in \cite{NijPutt}. To complement those solutions with the ones for the auxiliary variables \eqref{eq:defs} we give:   
\bse\begin{align}
v_a &=1-\boldsymbol{c}\cdot(\boldsymbol{I}+\boldsymbol{M})^{-1}\cdot(a\boldsymbol{I}+\boldsymbol{k})^{-1}\cdot\boldsymbol{r} \ , \\ 
s_a &=\boldsymbol{c}\cdot(\boldsymbol{I}+
\boldsymbol{M})^{-1}\cdot(a\boldsymbol{I}+\boldsymbol{k})^{-1}\cdot\boldsymbol{K}^{-1}\cdot\boldsymbol{r} \\
w_a &=\boldsymbol{c}\cdot\boldsymbol{K}^{-1}\cdot(\boldsymbol{I}+\boldsymbol{M})^{-1}
\cdot(a\boldsymbol{I}+\boldsymbol{k})^{-1}\cdot \boldsymbol{r}\ , \\  
h_a &=1+g\,\boldsymbol{c}\cdot\boldsymbol{K}^{-1}\cdot(\boldsymbol{I}+\boldsymbol{M})^{-1}\cdot(a\boldsymbol{I}+\boldsymbol{k})^{-1}\cdot\boldsymbol{K}^{-1}\cdot\boldsymbol{r}\ , 
\end{align}\ese 
where $a$~ is an arbitrary auxiliary parameter, as well as for the variables \eqref{eq:pqrt} 
\bse\begin{align}
r_a &=a-\boldsymbol{c}\cdot\boldsymbol{k}\cdot 
(\boldsymbol{I}+\boldsymbol{M})^{-1}\cdot(a\boldsymbol{I}+\boldsymbol{k})^{-1}\cdot\boldsymbol{r} \ , \\ 
q_a &=\boldsymbol{c}\cdot\boldsymbol{k}\cdot(\boldsymbol{I}+
\boldsymbol{M})^{-1}\cdot(a\boldsymbol{I}+\boldsymbol{k})^{-1}\cdot\boldsymbol{K}^{-1}\cdot\boldsymbol{r} \\
p_a &=\boldsymbol{c}\cdot\boldsymbol{K}^{-1}\cdot\boldsymbol{k}\cdot(\boldsymbol{I}+\boldsymbol{M})^{-1}
\cdot(a\boldsymbol{I}+\boldsymbol{k})^{-1}\cdot \boldsymbol{r}\ , \\  
t_a &=a+g\,\boldsymbol{c}\cdot\boldsymbol{K}^{-1}\cdot\boldsymbol{k}\cdot(\boldsymbol{I}+\boldsymbol{M})^{-1}\cdot(a\boldsymbol{I}+\boldsymbol{k})^{-1}\cdot\boldsymbol{K}^{-1}\cdot\boldsymbol{r}\ , 
\end{align}\ese 
and the 2-index variables \eqref{eq:defs2} 
\bse\begin{align}
s_{a,b} &=\boldsymbol{c}\cdot(b\boldsymbol{I}+\boldsymbol{k})^{-1}\cdot(\boldsymbol{I}+\boldsymbol{M})^{-1}\cdot(a\boldsymbol{I}+\boldsymbol{k})^{-1}\cdot\boldsymbol{r} \ , \\ 
t_{a,b} &=\boldsymbol{c}\cdot(b\boldsymbol{I}+\boldsymbol{k})^{-1}\cdot(\boldsymbol{I}+
\boldsymbol{M})^{-1}\cdot(a\boldsymbol{I}+\boldsymbol{k})^{-1}\cdot\boldsymbol{K}^{-1}\cdot\boldsymbol{r} \\
%w_a &=\boldsymbol{c}\cdot\boldsymbol{K}^{-1}\cdot(a\boldsymbol{I}+\boldsymbol{k})^{-1}
%\cdot(\boldsymbol{I}+\boldsymbol{M})^{-1}
%\cdot(a\boldsymbol{I}+\boldsymbol{k})^{-1}\cdot \boldsymbol{r}\ , \\  
u_{a,b} &=g\,\boldsymbol{c}\cdot\boldsymbol{K}^{-1}\cdot(b\boldsymbol{I}+\boldsymbol{k})^{-1}\cdot(\boldsymbol{I}+\boldsymbol{M})^{-1}\cdot(a\boldsymbol{I}+\boldsymbol{k})^{-1}\cdot\boldsymbol{K}^{-1}\cdot\boldsymbol{r}\ . 
\end{align}\ese 
Here we used the notations:
\[ \boldsymbol{k}={\rm diag}(k_1,\cdots,k_N)\ , \quad \boldsymbol{K}={\rm diag}(K_1,\cdots,K_N)\ ,\quad \boldsymbol{r}=(r_1,\cdots,r_N)^T\ , 
\]
and where $\boldsymbol{c}=(c_1,\cdots,c_N)$ is the vector of constants $c_j$. 
Then, the basic relations for the $\tau$-function 
can be expressed as 
\[ \frac{\wt{\tau}}{\tau}= v_{-p}h_{-p}-gw_{-p}s_{-p}\quad {\rm and}\quad  
\frac{\pl}{\pl p}\ln\,\tau=n(s_{p,-p}-u_{p,-p})\ , \]
and similar relations for the other shift and derivatives w.r.t. $q$.  
An open problem is to find a closed-form (possibly bilinear) elliptic 
difference equation for the $\tau$-function. Unlike in the non-elliptic 
($g=e=0$) case the basic relations are in only 
terms of 2$\times$ 2 matrix, while we need relations for the matrices 
themselves to derive such equations. How to supplement these is still 
an open problem.

\section{Elliptic Generating PDE} 
\setcounter{equation}{0}

The lattice systems which are derived from the elliptic matrix relations of 
the type \eqref{eq:Udynscont1} are compatible with systems of continuous 
equations derived from \eqref{eq:Udynscont1}, in which the discrete 
variables $n_\nu$ appear as parameters and where the continuous variables are 
the lattice parameters of the discrete equations. This role reversal between 
parameters and variables is a fundamental feature that manifests the duallity 
between integrable discrete and continuous systems, which goes deeper than 
the latter being just continuum limits (i.e. degenerations) of the former. 
Furthermore, the vector fields associated with the continuous flows 
are in fact (master) symmetries of the discrete equations. In this section we 
will derive a fundamental system of PDEs for the quantities defined in the 
previous sections where the independent variables are the lattice parameters 
$p_\nu$. These equations are non-autonomous, whereas the autonomous 
versions are the symmetries of the those same equations, but instead in terms 
of so-called \textit{Miwa variables}\footnote{Miwa variables are variables that 
encode the entire integrable hierarchy of PDEs generated by higher symmetries, i.e. they are given by vector fields  
\[ \frac{\pl}{\pl\xi_p}=\sum_{j=0}^\infty \frac{1}{p^{j+1}}\frac{\pl}{\pl t_j}\ , 
\quad \frac{\pl}{\pl\xi_q}=\sum_{j=0}^\infty \frac{1}{q^{j+1}}\frac{\pl}{\pl t_j}\ ,
\]
where $p$ and $q$ are lattice parameters, and the $t_j$ are the higher times of 
the associated continuous hierarchies. However, in this section the derivatives 
are with respect to the actual lattice parameters themselves. }. 

\subsection{Fundamental differential relations}

Starting from the fundamental 
continuous relations (\ref{eq:Udynscont1}) we can derive 
now a set of partial differential equations which are compatible 
with the discrete equations. The basic relations which can be derived from the 
definitions \eqref{eq:defs}, \eqref{eq:defs2} and \eqref{eq:pqrt} in combination 
with \eqref{eq:Ualg} will be collected in 2$\times$2 matrix forms that can be 
extracted from those relations, the first of which reads: 
\begin{equation}\label{eq:contS}
\frac{\pl}{\pl p}\bS =n\left(\boldsymbol{1}- G\,\bV_{-p}^T\,G^{-1}\,\bV_p\right) 
=n\left(\boldsymbol{1}- G\,\bV_{p}^T\,G^{-1}\,\bV_{-p}\right)\ , 
%=n\left[\boldmath{1}-2p(\,2p\boldmath{1}+\ut{\bS}-\wt{\bS}\,)^{-1}\right]\   , 
\end{equation}
(and a similar relation for $\pl\bS/\pl q$) where the matrix $\bS$ is given in (\ref{eq:Smat}). % and $G={\rm diag}(1,-g)$. 

To proceed further it is useful to introduce two more 2$\times$2 matrices, which 
are composed from entries of both $\bV_a$ and $\bR_a$, namely 
\begin{equation}\label{eq:RQ}  
\bW_a\equiv\left(\begin{array}{cc} 
w_a & -p_a \\ h_a & -t_a \end{array}\right)\qquad ,\qquad 
\bQ_a\equiv\left(\begin{array}{cc} 
-q_a & s_a \\ -r_a & v_a \end{array}\right) \  , 
\end{equation} 
which can be expressed in terms of the earlier matrices as 
follows: 
\bse\label{eq:matrels}\begin{eqnarray}
\left(\begin{array}{cc} 1&0\\0&-1\end{array}\right) \bW_a&=&
\bR_a\left(\begin{array}{cc} 0&0\\0&1\end{array}\right) - 
\bV_a\left(\begin{array}{cc} 0&0\\1&0\end{array}\right) \\ 
\left(\begin{array}{cc} 0&1\\g&0\end{array}\right) \bQ_a&=&
\bV_a\left(\begin{array}{cc} 0&1\\0&0\end{array}\right) - 
\bR_a\left(\begin{array}{cc} 1&0\\0&0\end{array}\right)\  .  
\end{eqnarray}\ese 
Furthermore, the following matrices depending on the auxiliary parameters $a$ 
and $b$ will come in handy: 
\begin{equation}\label{eq:ABmat}
\bA\equiv\left(\begin{array}{cc} 
1 & 0 \\ a^2-3e & 1 \end{array}\right)\qquad ,\qquad  
\bB\equiv\left(\begin{array}{cc} 
1 & 0 \\ b^2-3e & 1 \end{array}\right)\   , 
\end{equation}
which allow us to write a relation between matrices $\bW$ and $\bQ$, arising 
from the relations \eqref{eq:pqrtrels} or \eqref{eq:vwshrels}, namely  
\begin{equation}\label{eq:RQ}  
%\bW_a\,\bG^{-1}=\bA\,\bQ_a 
\bW_a=\bA\,\bQ_a\,\bG\quad  , \quad 
\bW_b=\bB\,\bQ_b\,\bG\  ,  
%\left(\begin{array}{cc} 
%w_a & -p_a \\ h_a & -t_a \end{array}\right)\boldsymbol{G}^{-1}=
%\left(\begin{array}{cc} 
%1 & 0 \\ a^2-3e & 1 \end{array}\right) 
%\left(\begin{array}{cc} 
%-q_a & s_a \\ -r_a & v_a \end{array}\right) \  , 
\end{equation} 
in which $\bG$ denotes the gauge matrix of \eqref{eq:gauge}. 

We are now in a position to complement the differential relation for the 
matrix $\bS$, \eqref{eq:contS}, with the 
following relations for the gauge matrix $\bG$ of \eqref{eq:gauge}, 
namely % from \eqref{eq:contmat} we have  
\bse\label{eq:contG}\begin{eqnarray} 
\frac{\pl\bG}{\pl p}&=& n\left[\cof(\bQ_p)\bsg_3\bW_{-p}+\left( 
\begin{array}{cc} 0 & -s \\ s & v-w\end{array}\right)\right] \   ,  
\label{eq:contGa} \\ 
\frac{\pl}{\pl p}\bG^{-1} &=& n\left[\cof(\bW_p)\bsg_3\bQ_{-p}+\left( 
\begin{array}{cc} w-v & -s \\ s & 0\end{array}\right) \right] \   ,  \label{eq:contGb} 
\end{eqnarray}\ese 
in which ~$\cof(\bW)=\det(\bW)\,\bW^{-1}$~ denotes the matrix of cofactors of a matrix $\bW$ and 
~$\bsg_3=\textrm{diag}(1,-1)$~ being the third Pauli matrix. The 
latter relation indicates that the gauge matrix $\bG$ in a sense interpolates between the matrices 
$\bQ$ and $\bW$, and in fact we have by applying \eqref{eq:RQ} the two following more specific relations:
\bse\label{eq:GG}\begin{eqnarray}
\left(\frac{\pl}{\pl p}\bG\right)\bG^{-1} &=& n\,\left[ \left(\begin{array}{cc} 0&0\\1&0\end{array}\right) + 
\left(\begin{array}{c} s\\w\end{array}\right) \left( -w,s\right) + \cof(\bQ_p)\bsg_3\bP\bQ_{-p}\right] \\  
\left(\frac{\pl}{\pl p}\bG^{-1}\right)\bG &=& n\,\left[ \left(\begin{array}{cc} 0&1\\0&0\end{array}\right) + 
\left(\begin{array}{c} -v\\s\end{array}\right) \left( s,v\right) + \cof(\bW_p)\bsg_3\bP^{-1}\bW_{-p}\right] \  ,   
\end{eqnarray}\ese 
in which the matrix $\bP$ is the one like \eqref{eq:ABmat} with $a$ or $b$ replaced by $p$. 

{\bf Remark:} We note the following relation between the determinants of the matrices 
$\bQ_a$ and $\bV_a$ 
%which can be obtained by exploiting eqs. \eqref{eq:vwhs} 
\begin{equation}\label{eq:Qcdet}
\det(\wt{\bQ}_a)=s\,\det(\wt{\bV}_a)+(p-a)\left( s_a\wt{v}_a-v_a\wt{s}_a\right)\  , 
\end{equation} 
leading (taking $a=p$) to the relation  
\begin{equation}\label{eq:Qdet}
\det(\wt{\bQ}_p)=s\,\det(\wt{\bV}_p)=s\frac{\tau}{\wt{\tau}}\quad \Leftrightarrow \quad 
\det(\bQ_{-p})=\wt{s}\frac{\wt{\tau}}{\tau}\  , 
\end{equation} 
which seems to indicate that the quantity $s$ could be regarded as a companion to the 
$\tau$-function. 

\subsection{Non-autonomous D$\Delta$-system:}

As an intermediate stage between fully discrete lattice equations and a 
fully continuous system of PDEs we mention that  
%From the relations \eqref{eq:contmat} in combination with the discrete relations 
%\eqref{eq:vwhs} %(and their 
%reverse relations obtained by interchanging $a\leftrightarrow -a$ and $%
%\wt{u}\leftrightarrow\underaccent{\wtilde}{u}$, 
%etc.), 
one can derive the following coupled systems of differential-difference equations. 
This relies on the following curious relations (which hold for all $a$): 
\bse\label{eq:strangeness}\begin{align}
& w_a v_{-a}-v_a w_{-a}=s_a h_{-a}-h_a s_{-a}\  , \\ 
& v_a h_{-a}-h_a v_{-a}+g( w_a s_{-a}-s_aw_{-a})+(a^2-3e)(s_a h_{-a}-h_a s_{-a})=2as\  ,
\end{align}\ese 
in the derivation, where the latter is a special case of \eqref{eq:strange} 
taking $b=-a$, and the former a consequence of the symmetry  $\,^{t\!}\bU=\bU$ when 
applied to \eqref{eq:Udynscont1}, with the latter multiplied from the left 
by $\bL^{-1}$. 

\iffalse 
\bse\label{eq:interDD}\bea 
\left(a-\wt{u}+\frac{w}{s}\right) \left( n-\frac{\pl u}{\pl a}\right) &=& 
\eea\ese 
\fi 

From the relations \eqref{eq:contS} and \eqref{eq:contG} in combination with 
\eqref{eq:RSmat} together with the identification \eqref{eq:Rrel} 
%\eqref{eq:vwhsb}, \eqref{eq:vwhsd} and \eqref{eq:vwhse} and \eqref{eq:strangeness}, as 
%well as \eqref{eq:contmatc} in combination with the Miura type relation %\eqref{eq:Miuraa}, 
we can derive the following non-autonomous differential-difference system: 
\bse\label{eq:fundDD}\begin{eqnarray}
&& (2p+\underaccent{\wtilde}{u}-\wt{u})\left(\frac{\pl u}{\pl p}-n\right) + 
2np+g(\wt{s}-\underaccent{\wtilde}{s})\frac{\pl s}{\pl p}=0\  , \label{eq:fundDDa} \\ 
&& \left[(p+\underaccent{\wtilde}{u})\frac{\pl s}{\pl p}+ns
-\frac{\pl w}{\pl p}\right]\wt{s}  
+ \left[(p-\wt{u})\frac{\pl s}{\pl p}-ns+\frac{\pl w}{\pl p}\right]\underaccent{\wtilde}{s}=0\  ,   
\label{eq:fundDDb} \\ 
&& \left[\left( p+u-\frac{\wt{w}}{\wt{s}}\right)\wt{s}+ \left( p-u+\frac{\underaccent{\wtilde}{w}}
{\underaccent{\wtilde}{s}}\right) \underaccent{\wtilde}{s}\right] \frac{\pl s}{\pl p}= 
s(\wt{s}-\underaccent{\wtilde}{s})\left(\frac{\pl u}{\pl p}-n\right)\  , 
\label{eq:fundDDc} \\
&& \left(p+u-\frac{\wt{w}}{\wt{s}}\right)\left(p-\wt{u}+
\frac{w}{s}\right)=p^2-\left(\frac{1}{s\wt{s}}+3e+g s\wt{s}\right)\  .  \label{eq:fundDDd} 
\end{eqnarray}\ese 
Following \cite{TTP} the vector field associated with this D$\Delta$ system 
can be viewed as a \textit{master symmetry} for the elliptic KdV system. Note that 
its autonomous form \eqref{eq:skewlim}, in a similar vein, corresponds to a 
hierarchy of symmetries of the EllKdV system in terms of a Miwa variable, as 
the corresponding time-variable. 
%A similar set of relations holds for the dual 
%system, but we omit the details.  

\subsection{Extended continuous matrix system} 

In order to complete the system of differential relations, 
we need also the ones for the matrices $\bV_a$ and $\bR_a$, which 
turn out to be, respectively 
\begin{equation}\label{eq:contV}
\frac{\pl}{\pl p}\bV_a=n\left[ \frac{1}{p-a}(\bV_a-\bV_p)
-\bZ_{a,-p}\,\bV_p\,\right]\ ,  
\end{equation}
and 
\begin{equation}\label{eq:contR}
\frac{\pl}{\pl p}\bR_a=n\left[ \bV_a+\frac{1}{p-a}(\bR_a-\bR_p) 
-\bZ_{a,-p}\,\bR_p\,\right]\   ,  
\end{equation}
while for the matrices $\bZ_{a,b}$ we have 
\begin{equation}\label{eq:contZ}
\frac{\pl}{\pl p}\bZ_{a,b}=n\left[ \frac{1}{p-a}(\bZ_{a,b}-\bZ_{p,b}) 
+\frac{1}{p+b}(\bZ_{a,-p}-\bZ_{a,b}) -\bZ_{a,-p}\,\bZ_{p,b}\,\right]\   .    
\end{equation}
%which follows from \eqref{eq:contmat3}. 
For the sake of completeness, we also mention the matrix differential relations 
for the matrices $\bW_a$ and $\bQ_a$, namely
\begin{equation}\label{eq:contW}
\frac{\pl}{\pl p}\bW_a=n\left[ \frac{1}{p-a}(\bW_a-\bW_p)-\bW_a\left(\begin{array}{cc} 
0&1\\0&0\end{array}\right)-\bsg_3\bZ_{a,-p}\bsg_3\bW_p\,\right]\   ,  
\end{equation}
and 
\begin{equation}\label{eq:contQ}
\frac{\pl}{\pl p}\bQ_a=n\left[ \frac{1}{p-a}(\bQ_a-\bQ_p)
-\bQ_a\left(\begin{array}{cc} 
0&0\\1&0\end{array}\right)-\cof(\bZ_{-p,a})\bQ_p\,\right]\  . 
\end{equation}
%by assembling the relations \eqref{eq:contmat1} and \eqref{eq:contmat2} in a different way. 
Finally, for the $\tau$-function we have: 
\begin{equation}\label{eq:conttau}
\frac{\pl}{\pl p}\ln\tau=n\,{\rm tr}(\bZ_{p,-p})\   , 
\end{equation}
which follows directly from \eqref{eq:taudiffrel}.

\subsection{Non-isospectral continuous Lax pair}  

The continuous relations can now be implemented, and mixed with the 
discrete lattice shift to obtain semi-discrete integrable equations. 
The main continuous relations for the main quantities $u$, $v$, $w$, $s$, 
$h$, are obtained from non-isospectral Lax pairs, which can be derived 
from the structure of subsection 3.6 setting 
\bea 
&& \vf=(p-k)^n(q-k)^m\left(\begin{array}{c} (\bu(\kp))_0 \\ (\bu(\kp))_{1}\end{array}\right)\  , 
\quad {\rm in\ \ which}\ \nn  \\ 
&& (\bu(\kp))_{2i}\equiv\tee\,\bL^i\,\bu(\kp)\ , \quad 
(\bu(\kp))_{2i+1}\equiv\tee\,\bL^i\,\bLd\,\bu(\kp)\ .  \nn 
%&& u_a(\kp)\equiv\tee\,(a+\Ld)^{-1}\cdot\bu(\kp)\,\bee\qquad ,\qquad 
%s_a(\kp)\equiv\tee\,(a+\Ld)^{-1}\cdot\bL^{-1}\cdot\bu(\kp)\,\bee\ .  \nn 
\eea
These linear problems take the form 
\begin{equation}\label{eq:cLax}
\frac{\pl\vf}{\pl p} = \left[ n\bF + \frac{\bP_0+\frac{g}{K}\bP_1}{k^2-p^2}\right]\vf \   , \quad  
\frac{\pl\vf}{\pl q} = \left[ m\bF + \frac{\bQ_0+\frac{g}{K}\bQ_1}{k^2-q^2}\right]\vf \   , 
\end{equation} 
in which $\bF=\left(\begin{array}{cc} 0 & 0\\ 1 & 0\end{array}\right)$ 
as before, and where the matrices $\bP_0$ and $\bP_1$ are given by 
\bse\begin{align} 
\bP_0 &=\left(\begin{array}{ccc} 
\frac{\pl U_{1,0}}{\pl p} -n(p+u) &,& n-\frac{\pl u}{\pl p} \\ 
\frac{\pl U_{1,1}}{\pl p} +n(p^2-2U_{1,0}) &,&  -\frac{\pl U_{1,0}}{\pl p} -n(p-u) \end{array}\right)  \\ 
\bP_1 & = 
 \left(\begin{array}{ccc} 
-s\frac{\pl U_{1,-1}}{\pl p} -v\frac{\pl w}{\pl p}+nsw 
&,& v\frac{\pl s}{\pl p} -s\frac{\pl v}{\pl p}-ns^2\\ 
U_{1,-1}\frac{\pl w}{\pl p}-w\frac{\pl U_{1,-1}}{\pl p} -n(1-w^2) &,& 
s\frac{\pl U_{1,-1}}{\pl p} +v\frac{\pl w}{\pl p}-nsw 
\end{array}\right)\vf\ ,  
\end{align}\ese 
and similar expressions for $\bQ_0$ and $\bQ_1$ replacing $p$ by $q$ and $n$ by $m$, where the latter now play the role of parameters. They contain 
the six dependent variables  $u$, $v$, $w$, $s$, $U_{1,0}$ and $U_{1,1}$, while 
$U_{1,-1}=(1-v w)/s$. 
%A dual system for the gauge-related quantity $\vr$ to $\vf$ 
%is given in Appendix D, which contains dependent variables $h$, $v$, $w$, $s$, $U_{-1,-1}$ and $U_{-1,-2}$ and which yield equations of a similar structure 
%as the system given below, but we will abstain from presenting it here. 

The consistency relation for \eqref{eq:cLax}, i.e. 
\[ \frac{\pl}{\pl p}\left(\frac{\pl\vf}{\pl q}\right) 
=\frac{\pl}{\pl q}\left(\frac{\pl\vf}{\pl p}\right)\   , \] 
yields the set of conditions 
\bse\label{eq:ccons}\begin{eqnarray} 
0&=&\frac{\pl\bP_0}{\pl q}+m\left[ \bP_0\,,\,\bF\,\right] + 
\frac{\left[ \bP_0\,,\,\bQ_0\,\right] -g \left[ \bP_1\,,\,\bQ_1\,\right]}{p^2-q^2}\   ,  \label{eq:cconsa}  \\  
0&=&\frac{\pl\bQ_0}{\pl p}+n\left[ \bQ_0\,,\,\bF\,\right] + 
\frac{\left[ \bP_0\,,\,\bQ_0\,\right] -g \left[ \bP_1\,,\,\bQ_1\,\right]}{p^2-q^2}\   ,  \label{eq:cconsb}  \\  
0&=&\frac{\pl\bP_1}{\pl q}+m\left[ \bP_1\,,\,\bF\,\right] 
+\frac{\left[ \bP_0\,,\,\bQ_1\,\right] + \left[ \bP_1\,,\,\bQ_0\,\right] 
+(p^2-3e) \left[ \bP_1\,,\,\bQ_1\,\right]}{p^2-q^2}\   ,  \label{eq:cconsc}  \\  
0&=&\frac{\pl\bQ_1}{\pl p}+n\left[ \bQ_1\,,\,\bF\,\right] 
+\frac{\left[ \bP_0\,,\,\bQ_1\,\right] + \left[ \bP_1\,,\,\bQ_0\,\right] 
+(q^2-3e) \left[ \bP_1\,,\,\bQ_1\,\right]}{p^2-q^2}\   .  \label{eq:cconsd}   
\end{eqnarray}\ese  
%In eqs. \eqref{eq:ccons} the matrices are obtained from 
%\be\label{eq:ABF}
%\bA_0=n\,{\rm cof}(\bQ_{-a})\,\left(\begin{array}{cc} 0&1\\ f&0\end{array}\right)\,\bQ_a\quad ,\quad 
%\bA_1=n\,{\rm cof}(\bQ_{-a})\,\bsg_3\,\bA\,\bQ_a\quad , 
%\ee 
%and similarly for $\bB_0$ and $\bB_1$~. 
Working out these consistency conditions we obtain on the one hand from the matrix entries of \eqref{eq:cconsa} the following set 
of coupled differential equations:
\bse\label{eq:PDElist1}\begin{eqnarray} 
&& (p^2-q^2)\frac{\pl^2 u}{\pl p\,\pl q} = 2\left\{ \left( \frac{\pl U_{1,0}}{\pl p} -nu\right) \left( m-\frac{\pl u}{\pl q}\right) 
- \left( \frac{\pl U_{1,0}}{\pl q} -mu\right) \left( n-\frac{\pl u}{\pl p}\right) \right\} \nn \\ 
&& -2g\left\{ \left( v\frac{\pl s}{\pl p}-s\frac{\pl v}{\pl p} -ns^2\right) 
\left( s\frac{\pl U_{1,-1}}{\pl q}+v\frac{\pl w}{\pl q}-msw\right) \right. \nn \\ 
&& \qquad \quad \left. -\left( v\frac{\pl s}{\pl q}-s\frac{\pl v}{\pl q} -ms^2\right) 
\left( s\frac{\pl U_{1,-1}}{\pl p}+v\frac{\pl w}{\pl p}-nsw\right) \right\}\  , \\ 
&& (p^2-q^2)\left( \frac{\pl^2 U_{1,0}}{\pl p\,\pl q} -m\frac{\pl u}{\pl p}-n\frac{\pl u}{\pl q}+mn\right) = \nn \\ 
&& = \left\{ \left(m-\frac{\pl u}{\pl q}\right)\left( \frac{\pl U_{1,1}}{\pl p} +n(p^2-2U_{1,0})\right) 
- \left(n-\frac{\pl u}{\pl p}\right)\left( \frac{\pl U_{1,1}}{\pl q} +m(q^2-2U_{1,0})\right) \right\} \nn \\ 
&& +g\left\{ \left( v\frac{\pl s}{\pl p}-s\frac{\pl v}{\pl p} -ns^2\right) 
\left( U_{1,-1}\frac{\pl w}{\pl q}-w\frac{\pl U_{1,-1}}{\pl q}+v\frac{\pl w}{\pl q}-m(1-w^2)w\right) \right. \nn \\ 
&& \qquad  \left. -\left( v\frac{\pl s}{\pl q}-s\frac{\pl v}{\pl q} -ms^2\right) 
\left( U_{1,-1}\frac{\pl w}{\pl p}-w\frac{\pl U_{1,-1}}{\pl p}+v\frac{\pl w}{\pl p}-n(1-w^2)\right) \right\}\  , \\ 
&& (p^2-q^2)\left( \frac{\pl^2 U_{1,1}}{\pl p\,\pl q} -2m\frac{\pl U_{1,0}}{\pl p}-2n\frac{\pl U_{1,0}}{\pl q}+2mnu\right) = \nn \\ 
&& = 2\left\{ \left(mu-\frac{\pl U_{1,0}}{\pl q}\right)\left( \frac{\pl U_{1,1}}{\pl p} +n(p^2-2U_{1,0})\right) 
- \left(nu-\frac{\pl U_{1,0}}{\pl p}\right)\left( \frac{\pl U_{1,1}}{\pl q} +m(q^2-2U_{1,0})\right) \right\} \nn \\ 
&& ~-2g\left\{ \left( s\frac{\pl U_{1,-1}}{\pl p}+v\frac{\pl w}{\pl p} -nsw\right) 
\left( U_{1,-1}\frac{\pl w}{\pl q}-w\frac{\pl U_{1,-1}}{\pl q}+v\frac{\pl w}{\pl q}-m(1-w^2)w\right) \right. \nn \\ 
&& \qquad  \left. -\left( s\frac{\pl U_{1,-1}}{\pl q}+v\frac{\pl w}{\pl q} -msw\right) 
\left( U_{1,-1}\frac{\pl w}{\pl p}-w\frac{\pl U_{1,-1}}{\pl p}+v\frac{\pl w}{\pl p}-n(1-w^2)\right) \right\}\  , 
\end{eqnarray}  \ese 
whilst the same set of equations follow from the matrix entries of \eqref{eq:cconsb}, (in fact eqs. \eqref{eq:PDElist1} are 
covariant under the interchange of the variables $p$ and $q$, and the parameters $n$ and $m$). On the other hand, from 
\eqref{eq:cconsc} we get: 
\bse\label{eq:PDElist2}\begin{eqnarray} 
&& (p^2-q^2)\left\{ \frac{\pl}{\pl q}\left( s\frac{\pl U_{1,1}}{\pl p} + v\frac{\pl w}{\pl p}-nsw\right) 
-m\left( v\frac{\pl s}{\pl p}-s\frac{\pl v}{\pl p}-ns^2\right)\right\} = \nn \\ 
&& = \left(n-\frac{\pl u}{\pl p}\right)\left( U_{1,-1}\frac{\pl w}{\pl q}-w\frac{\pl U_{1,-1}}{\pl q} -m(1-w^2)\right)  
- \left(m-\frac{\pl u}{\pl q}\right)\left( U_{1,-1}\frac{\pl w}{\pl p}-w\frac{\pl U_{1,-1}}{\pl p} -n(1-w^2)\right) \nn \\ 
&& + \left( v\frac{\pl s}{\pl p}-s\frac{\pl v}{\pl p} -ns^2\right) \left( \frac{\pl U_{1,1}}{\pl q} +m(q^2-2U_{1,0})\right) 
- \left( v\frac{\pl s}{\pl q}-s\frac{\pl v}{\pl q} -ms^2\right) \left( \frac{\pl U_{1,1}}{\pl p} +n(p^2-2U_{1,0})\right) \nn \\ 
&& -(p^2-3e)\left\{ \left( U_{1,-1}\frac{\pl w}{\pl p}-w\frac{\pl U_{1,-1}}{\pl p} -n(1-w^2)\right) 
\left( v\frac{\pl s}{\pl q}-s\frac{\pl v}{\pl q} -ms^2\right) \right. \nn \\ 
&& \qquad \qquad \qquad -\left. \left( U_{1,-1}\frac{\pl w}{\pl q}-w\frac{\pl U_{1,-1}}{\pl q}-m(1-w^2)\right) 
 \left( v\frac{\pl s}{\pl p}-s\frac{\pl v}{\pl p} -ns^2\right) \right\}\   ,  \\ 
&& (p^2-q^2)\frac{\pl}{\pl q}\left( v\frac{\pl s}{\pl p}-s\frac{\pl v}{\pl p}-ns^2\right)= 2\left\{ 
\left( s\frac{\pl U_{1,-1}}{\pl p}+v\frac{\pl w}{\pl p}-nsw\right) \left( m-\frac{\pl u}{\pl q}\right) \right. \nn \\ 
&& \qquad - \left( s\frac{\pl U_{1,-1}}{\pl q} +v\frac{\pl w}{\pl q} - msw\right) \left( n-\frac{\pl u}{\pl p}\right) 
+ \left( v\frac{\pl s}{\pl p}-s\frac{\pl v}{\pl p} -ns^2\right) \left( \frac{\pl U_{1,0}}{\pl q} -mu\right) \nn \\ 
&& \qquad \qquad \left. - \left( v\frac{\pl s}{\pl q}-s\frac{\pl v}{\pl q} -ms^2\right) \left( \frac{\pl U_{1,0}}{\pl p} -nu\right) 
\right\}  \nn \\ 
&& \qquad +2(p^2-3e) \left\{ \left( s\frac{\pl U_{1,-1}}{\pl p}+v\frac{\pl w}{\pl p}-nsw\right) 
\left( v\frac{\pl s}{\pl q} -s\frac{\pl v}{\pl q}-ms^2\right) \right. \nn \\ 
&& \qquad \qquad \qquad \qquad \left. -\left( s\frac{\pl U_{1,-1}}{\pl q}+v\frac{\pl w}{\pl q}-msw\right) 
\left( v\frac{\pl s}{\pl p} -s\frac{\pl v}{\pl p}-ns^2\right) \right\} \   ,  \\ 
&& (p^2-q^2)\left\{ \frac{\pl}{\pl q}\left( U_{1,-1}\frac{\pl w}{\pl p} -w\frac{\pl U_{1,-1}}{\pl p}-n(1-w^2)\right) 
+2m\left( s\frac{\pl U_{1,-1}}{\pl p}+v\frac{\pl w}{\pl p}-nsw\right)\right\} = \nn \\ 
&& = 2\left\{ \left( \frac{\pl U_{1,1}}{\pl p}+n(p^2-2U_{1,0})\right)\left( s\frac{\pl U_{1,-1}}{\pl q}+v\frac{\pl w}{\pl q} -msw\right) \right. \nn \\ 
&& \qquad\qquad - \left( \frac{\pl U_{1,1}}{\pl q} +m(q^2-2U_{1,0})\right) \left( s\frac{\pl U_{1,-1}}{\pl p}+v\frac{\pl w}{\pl p} -nsw\right) \nn \\ 
&& \qquad - \left( U_{1,-1}\frac{\pl w}{\pl p} -w\frac{\pl U_{1,-1}}{\pl p}-n(1-w^2)\right) \left( \frac{\pl U_{1,0}}{\pl q}-mu\right) \nn \\ 
&& \qquad \qquad + \left.\left( U_{1,-1}\frac{\pl w}{\pl q} -w\frac{\pl U_{1,-1}}{\pl q}-m(1-w^2)\right) \left( \frac{\pl U_{1,0}}{\pl p}-nu\right) \right\} \nn \\ 
&& +2(p^2-3e)\left\{ \left( U_{1,-1}\frac{\pl w}{\pl p} -w\frac{\pl U_{1,-1}}{\pl p}-n(1-w^2)\right) 
\left( s\frac{\pl U_{1,-1}}{\pl q}+v\frac{\pl w}{\pl q}-msw\right) \right. \nn \\ 
&& \qquad\qquad \left. - \left( U_{1,-1}\frac{\pl w}{\pl q} -w\frac{\pl U_{1,-1}}{\pl q}-m(1-w^2)\right) 
\left( s\frac{\pl U_{1,-1}}{\pl p}+v\frac{\pl w}{\pl p}-nsw\right) \right\} \   .   
\end{eqnarray}\ese  
One can show, using the fact that ~$U_{1,-1}=(1-vw)/s$~, that eqs. \eqref{eq:PDElist2}, albeit seemingly asymmetric w.r.t. the interchange of 
the variables $p$ and $q$ and parameters $n$ and $m$, are still covariant, and hence we obtain the same set of equations from 
\eqref{eq:cconsd}. Thus, the six second-order PDEs consisting of \eqref{eq:PDElist1} together with \eqref{eq:PDElist2} forms the full list of 
equations that we obtain from the compatibility conditions \eqref{eq:ccons}. In fact, they constitute a closed-form coupled set of PDEs for the six 
quantities $u$, $s$, $v$, $w$, $U_{1,0}$ and $U_{1,1}$. 

We note that setting $g=0$ in \eqref{eq:PDElist1} we get a coupled system 
between only the three quantities $u$, $U_{1,0}$ and $U_{1,1}$. This system 
is equivalent to the \textit{generating PDE} for the KdV hierarchy that was 
introduced in \cite{NHJ}, and which subsequently was shown in \cite{TTX} to 
constitute a generalisation of the Ernst equations of General Relativity. Thus, the full 
elliptic case comprising \eqref{eq:PDElist2} as well as \eqref{eq:PDElist1} 
could be viewed as an elliptic version of the Ernst equations.

\section{Discussion} 

This rather curious elliptic KdV system \eqref{eq:EllDDKdV} is  
nonetheless by construction a natural 
generalisation of some well-known quadrilateral lattice equations 
(the lattice KdV, modified and Schwarzian KdV equations), which exhibits some surprising new features. In fact, the pivotal 
quantity $s$ connecting the various parts of the system has not 
made an appearance before in the solution structure of the ABS lattice 
equations through direct linearisation (cf. \cite{AtkNij2010,NijAtk2010,AtkNijHiet2009,HietZh2009}). While many of the 
integrability aspects are still present in the EllKdV system (such as the MDC, Lax pairs, etc.) some of those exhibit new features compared to the non-elliptic case. For instance, 
the Lax pair no longer possesses a natural factorisation (as in the case 
of H1), the simplest nontrivial periodic reductions (in the sense of \cite{PNC}) are of higher-genus, the $\tau$-function representation 
seems no longer to be of bilinear form and the natural 
Lagrangian description fails. These lacunae are challenges to our current understanding of quad 
equations rather than negative features. 
While we established here some novel results on the EllKdV system, there remain still many tantalizing open problems: establish an underlying $r$-matrix structure, constructing algebro-geometric solutions or other solutions beyond the soliton ones, as well as characterising the intrinsic nature of the special reduction of the 2$\times$2 matrix lattice KdV systems in terms of some loop group. Furthermore, the connection between the 
EllKdV system, and its KP generalisation, cf. \cite{JennNij2014}, and 
the KP systems that are associated with the Landau-Lifschitz class, cf. 
\cite{Date,FuNij2022} associated with a `reverse' Cauchy kernel, 
exhibiting Pfaffian structure, remains open. 
% We aim to address these issues in the near future. 

\subsection*{Acknowledgement} 

Some preliminary results reported in this article appeared in a very preliminary
form in \cite{Puttock}, which remained unpublished.  FWN was supported by the Foreign Expert 
Program of the Ministry of Sciences and Technology of China, grant number G2023013065L. 
The project was supported by NSFC (Nos. 12171306, 12271334 and 12326428). \\ 
\vspace{.05cm} 

\noindent 
{\bf Conflict of interest:} 
The authors declare that they have no conflicts of interest.

\end{document}